\documentclass{article} 
\usepackage{iclr2025_conference,times}
\iclrfinalcopy

\usepackage{amsmath,amsfonts,bm}

\def\eqref#1{equation~\ref{#1}}

\def\1{\bm{1}}

\DeclareMathAlphabet{\mathsfit}{\encodingdefault}{\sfdefault}{m}{sl}
\SetMathAlphabet{\mathsfit}{bold}{\encodingdefault}{\sfdefault}{bx}{n}

\usepackage{wrapfig}
\usepackage{hyperref}
\usepackage{url}
\usepackage[most]{tcolorbox}
\usepackage{xspace}
\usepackage{longtable}
\usepackage{pifont}
\usepackage{times}
\usepackage{latexsym}
\usepackage{booktabs}
\usepackage{multirow}
\usepackage{graphicx}
\usepackage{enumitem}
\usepackage{algorithm}
\usepackage{algorithmic}
\usepackage{amsmath}
\usepackage{colortbl}
\usepackage{inconsolata}
\usepackage{enumitem}
\usepackage{pifont}
\usepackage[most]{tcolorbox}
\usepackage{xspace}
\usepackage{colortbl}
\usepackage{makecell}
\usepackage[utf8]{inputenc}
\usepackage{listings}
\usepackage{fancyvrb}
\newcommand{\methodnamews}{\textsc{HyperAgent}}

\newcommand{\methodname}{\methodnamews~}

\definecolor{lightgray}{RGB}{240,240,240}
\definecolor{headcolor}{RGB}{230,230,230}

\definecolor{swecream}{RGB}{255,247,236}

\definecolor{issueborder}{HTML}{15071A}
\definecolor{issuefill}{HTML}{F6F8FA}
\definecolor{envfill}{HTML}{FFF7F3}
\definecolor{envborder}{HTML}{D35B27}
\definecolor{agentfill}{HTML}{FBFCFE}
\definecolor{agentborder}{HTML}{0C69DA}
\definecolor{goldpatchborder}{HTML}{FABB00}
\definecolor{goldpatchfill}{HTML}{FFF7E1}

\DefineVerbatimEnvironment{CodeVerbatim}{Verbatim}{
 formatcom={\color{black}},
 fontsize=\small,
 fontfamily=\ttdefault,
 fontseries=\mddefault,
 fontshape=\updefault,
 fillcolor=\color{white},
 framerule=0pt,
}

\definecolor{lightgray}{RGB}{240,240,240}
\definecolor{headcolor}{RGB}{230,230,230}

\definecolor{swecream}{RGB}{255,247,236}

\definecolor{issueborder}{HTML}{15071A}
\definecolor{issuefill}{HTML}{F6F8FA}
\definecolor{envfill}{HTML}{FFF7F3}
\definecolor{envborder}{HTML}{D35B27}
\definecolor{agentfill}{HTML}{FBFCFE}
\definecolor{agentborder}{HTML}{0C69DA}
\definecolor{goldpatchborder}{HTML}{FABB00}
\definecolor{goldpatchfill}{HTML}{FFF7E1}

\newtcolorbox{observationbox}[1][]{
        colback=envfill,
        colbacktitle=envfill,
        colframe=envborder,
        arc=5pt,
        fontupper=\small,
        fonttitle=\bfseries\color{black},
        boxrule=0.5mm,
        boxsep=1mm,
        width=\linewidth,
        breakable,
        title={Observation \hfill #1},
        rounded corners,
        toptitle=0.7mm,
        bottomtitle=0.7mm
}
\newtcolorbox{goldpatchbox}[1][]{
        colback=goldpatchfill,
        colbacktitle=goldpatchfill,
        colframe=goldpatchborder,
        arc=5pt,
        fontupper=\small,
        fonttitle=\bfseries\color{black},
        boxrule=0.5mm,
        boxsep=1mm,
        width=\linewidth,
        breakable,
        title={Patch \hfill #1},
        rounded corners,
        toptitle=0.7mm,
        bottomtitle=0.7mm
}
\newtcolorbox{issuebox}[1][]{
        colback=issuefill,
        colbacktitle=issuefill,
        colframe=issueborder,
        arc=5pt,
        fontupper=\small,
        fonttitle=\bfseries\color{black},
        boxrule=0.5mm,
        boxsep=1mm,
        width=\linewidth,
        breakable,
        title={Issue \hfill #1},
        rounded corners,
        toptitle=1mm
}
\newtcolorbox{agentbox}[1][]{
        colback=agentfill,
        colbacktitle=agentfill,
        colframe=agentborder,
        arc=5pt,
        fontupper=\small,
        fonttitle=\bfseries\color{black},
        boxrule=0.5mm,
        boxsep=1mm,
        width=\linewidth,
        breakable,
        title={\methodname \hfill #1},
        rounded corners,
        toptitle=1mm,
        lower separated=false
}
\newtcolorbox{fileviewerbox}[1]{
        enhanced,
        breakable,
        boxrule = 1.5pt,
        fontupper = \small,
        fonttitle = \bf\color{black},
        arc = 5pt,
        rounded corners,
        colframe = black,
        colbacktitle = swecream,
        colback = swecream,
        title = #1,
        left=4pt 
}
\newtcolorbox{promptbox}[1]{
    enhanced,
    breakable,
    boxrule=1pt,  
    fontupper=\small,
    fonttitle=\bfseries\color{black},
    arc=3pt,  
    rounded corners,
    colframe=black,
    colbacktitle=swecream,
    colback=swecream,
    title=#1,
    left=2mm,  
    right=2mm,  
    top=1mm,  
    bottom=1mm  
}
\title{HyperAgent: Generalist Software Engineering Agents to Solve Coding Tasks at Scale}

\author{Huy N. Phan$^{\spadesuit}$ \quad 
	Tien N. Nguyen$^{\diamondsuit}$ \quad 
	Phong X. Nguyen$^{\spadesuit}$ \quad 
	Nghi D. Q. Bui$^{\spadesuit\heartsuit}$\thanks{Corresponding author: bdqnghi@gmail.com} \\
	$^{\spadesuit}$FPT Software AI Center, Viet Nam \\
	$^{\diamondsuit}$The University of Texas at Dallas, USA \\
	$^{\heartsuit}$\textit{Project Lead} }

\begin{document}

\maketitle

\begin{abstract}
    Large Language Models (LLMs) have transformed software engineering (SE), exhibiting exceptional abilities in various coding tasks. Although recent advancements have led to the development of autonomous software agents using LLMs for end-to-end development tasks, these systems are often tailored to specific SE tasks. We present \methodname, a novel generalist multi-agent system that addresses a broad spectrum of SE tasks across multiple programming languages by emulating the workflows of human developers. \methodname consists of four specialized agents—Planner, Navigator, Code Editor, and Executor—capable of managing the full lifecycle of SE tasks, from initial planning to final verification. \methodname achieves state-of-the-art results on diverse SE tasks, including GitHub issue resolution on the well-known SWE-Bench benchmark, surpassing strong baselines. Additionally, \methodname excels in repository-level code generation (RepoExec) and fault localization and program repair (Defects4J), frequently outperforming SOTA baselines.

\end{abstract}
\begin{center}
        \noindent \includegraphics[height=12pt]{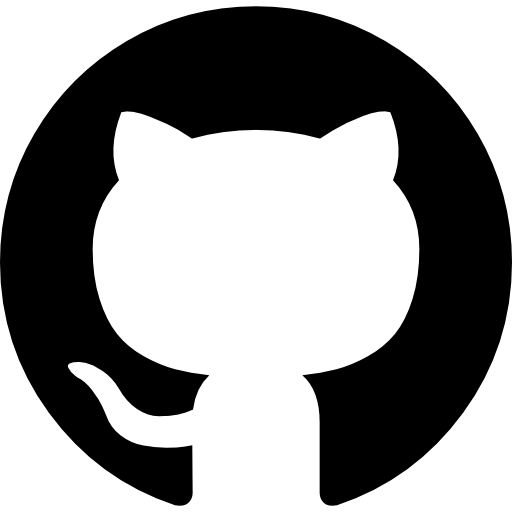} \quad \textbf{GitHub:} \href{https://github.com/FSoft-AI4Code/HyperAgent}{https://github.com/FSoft-AI4Code/HyperAgent}
\end{center}
\section{Introduction}
In recent years, Large Language Models (LLMs) have demonstrated remarkable capabilities in assisting with various coding tasks, ranging from code generation and completion to bug fixing and refactoring. These models have transformed the way developers interact with code, providing powerful tools that can understand and generate human-like code snippets with impressive accuracy. However, as software engineering tasks grow in complexity, there is an emerging need for more sophisticated solutions that can handle the intricacies of real-world software development.

\noindent Software agents built on LLMs have emerged as a promising solution to automate complex software engineering tasks, leveraging the advanced reasoning and generative abilities of LLMs. These agents can handle tasks such as code generation, bug localization, and orchestrating multi-step development processes. However, most current agents are limited in scope, typically focused on a \textbf{specific SE task}, such as resolving GitHub issues~\citep{jimenez2023swe, chen2024coder, arora2024masai, xia2024agentless, zhang2024diversity, yang2024swe} using benchmarks like SWE-bench~\citep{jimenez2023swe}, or tackling competitive code generation tasks like APPS~\citep{hendrycks2021measuring}, HumanEval~\citep{chen2021evaluating}, and MBPP~\citep{austin2021program}. Other agents~\citep{qian2024chatdev, hong2023metagpt, nguyen2024agilecoder} focus on generating complex software based on requirements. While these specialized agents excel in their domains, their claim of addressing general software engineering tasks is often overstated, as real-world SE challenges require more versatility across tasks, languages, and development scenarios.

\begin{figure*}[h]
	\centering
	\includegraphics[width=\textwidth]{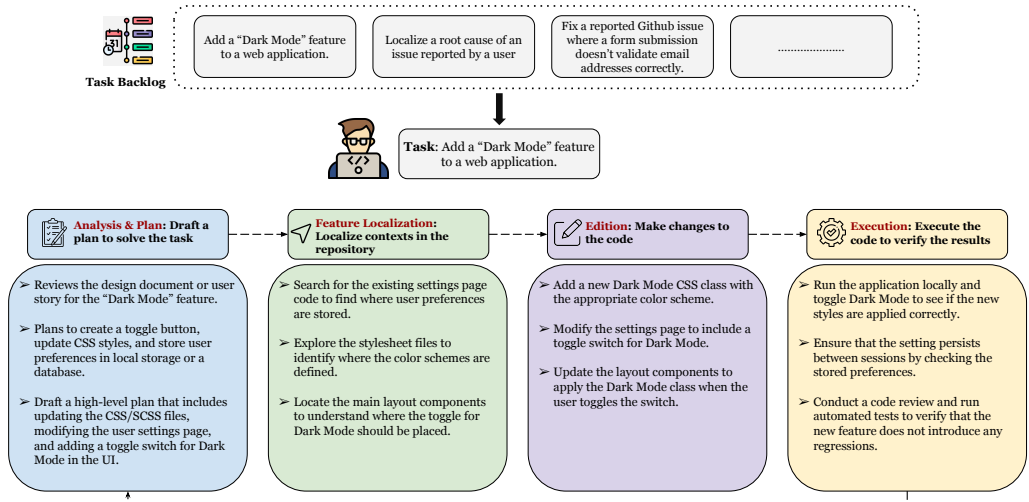}
	
	\caption{Illustration of a Developer’s Workflow for Resolving a Software Engineering Task. The diagram outlines the key phases a developer typically follows when implementing a new feature, such as adding a “Dark Mode” to a web application.}
	\label{fig:developer-workflow}
\end{figure*}

\noindent To address such drawbacks, we propose \methodnamews, a generalist multi-agent system designed to resolve a broad spectrum of SE tasks. Our design philosophy is rooted in the workflows that software engineers typically follow in their daily routines—whether it’s implementing new features in an existing codebase, localizing bugs in a large project, or providing fixes for reported issues and so on. While developers may use different tools or approaches to tackle these tasks, they generally adhere to consistent workflow patterns. We illustrate this concept through a workflow that represents how developers typically resolve coding tasks. Although different SE tasks require varied approaches, they all follow a similar workflow. 

Figure~\ref{fig:developer-workflow} illustrates a typical workflow for a software engineer when resolving a task from the backlog, which is a list of tasks to be completed within a specific period.




\begin{enumerate}[leftmargin=*]
    \item \textbf{Analysis \& Plan:} The developer starts by understanding the task requirements through documentation review and stakeholder discussions. A working plan is then formulated, outlining key steps, potential challenges, and expected outcomes. This plan remains flexible, adjusting as new insights are gained or challenges arise. 
    
    \item \textbf{Feature Localization:} With a plan in place, the developer navigates the \textit{repository} to identify relevant components, known as feature localization~\citep{michelon2021spectrum, martinez2018feature, castro2019pangolin}. This involves locating classes, functions, libraries, or modules pertinent to the task. Understanding dependencies and the system's overall design is crucial to make informed decisions later.
    
    \item \textbf{Edition:} The developer edits the identified code components, implementing changes or adding new functionality. This phase also involves ensuring smooth integration with the existing codebase, maintaining code quality, and adhering to best practices. 
    
    \item \textbf{Execution:} After editing, the developer tests the modified code to verify it meets the plan’s requirements. This includes running unit and integration tests, as well as conducting manual testing or peer reviews. If issues are found, the process loops back to previous phases until the task is fully resolved.
\end{enumerate}
These four steps are repeated until the developer confirms task completion. The exact process may vary depending on the task and the developer’s skill level; some tasks are completed in one phase, while others require multiple iterations—if the developer is unsatisfied after the Execution step, the entire process may repeat. In \methodnamews, the framework is organized around four primary agents: \textit{Planner}, \textit{Navigator}, \textit{Code Editor}, and \textit{Executor}, as illustrated in Figure~\ref{fig:main-fig}. Each agent corresponds to a specific step in the workflow shown in Figure~\ref{fig:developer-workflow}, though their workflows may differ slightly from how a human developer might approach similar tasks.\footnote{Details about each agent, along with how these advantages are achieved, are provided in Sections~\ref{sec:implementation}}. Our design emphasizes three main advantages over existing methods:
(1) Generalizability, the framework adapts easily to various tasks with minimal configuration, requiring little additional effort to incorporate new modules, (2) Efficiency, agents are optimized for processes with varying complexity, employing lightweight LLMs for tasks like navigation and more advanced models for code editing and execution and (3) Scalability, the system scales effectively in real-world scenarios with numerous subtasks, handling complex tasks efficiently.

Experimental results (See Section \ref{sec:eval}) highlight \methodname's unique position as the first system capable of working off-the-shelf across diverse software engineering tasks and programming languages, often exceeding specialized systems' performance. Its versatility positions \methodname as a transformative tool for real-world software development. In summary, the key contributions of this work include:

\begin{itemize}[leftmargin=*]
\item Introduction of \methodname, a generalist multi-agent system that closely mimics typical software engineering workflows and is able to handle a broad spectrum of software engineering tasks across different programming languages.
\item Extensive evaluation demonstrating superior performance across various software engineering benchmarks, including Github issue resolution (SWE-Bench-Python), repository-level code generation (RepoExec-Python), and fault localization and program repair (Defects4J-Java). To our knowledge, \methodname is the first system designed to work off-the-shelf across diverse SE tasks in multiple programming languages without task-specific adaptations.
\item Insights into the design and implementation of scalable, efficient, and generalizable software engineering agent systems, paving the way for more versatile AI-assisted development tools that can seamlessly integrate into various stages of the software lifecycle.
\end{itemize}

\section{Related Work}


\subsection{Deep Learning for Automated Programming}
In recent years, applying deep learning to automated programming has captured significant interest within the research community~\citep{balog2016deepcoder, bui2018hierarchical, bui2021treecaps, feng2020codebert, wang2021codet5, allamanis2018survey, bui2023codetf, guo2020graphcodebert, guo2022unixcoder}. Specifically, Code Large Language Models (CodeLLMs) have emerged as a specialized branch of LLMs, fine-tuned for programming tasks~\citep{wang2021codet5,wang2023codet5+, feng2020codebert, allal2023santacoder, li2023starcoder, lozhkov2024starcoder, guo2024deepseek, pinnaparaju2024stable, zheng2024opencodeinterpreter, roziere2023code, nijkamp2022codegen, luo2023wizardcoder, xu2022systematic, bui2022detect}. These models have become foundational in building AI-assisted tools for developers, aiming to solve competitive coding problems from benchmarks such as HumanEval \citep{chen2021humaneval}, MBPP \citep{austin2021program}, APPs~\citep{hendrycks2021measuring} and CRUXEval~\cite{gu2024cruxeval}.

\subsection{Benchmarks for Software Engineering}

Recent works have introduced SE benchmarks that expand evaluation criteria by incorporating third-party libraries~\citep{lai2023ds,liu2023ml}, derivative code completion tasks~\citep{muennighoff2023octopack}, test coverage~\citep{liu2023your}, modified edit scope~\citep{ding2024crosscodeeval,yu2024codereval,du2023classeval}, and robustness to dataset contamination~\citep{naman2024livecodebench}. However, these benchmarks often remain limited to short, self-contained code problems, typically requiring basic language primitives. As LMs advance, many benchmarks are becoming saturated, prompting a need for more complex tasks involving deeper reasoning and problem-solving. Efforts like SWE-bench~\citep{jimenez2023swe} simulate GitHub issue resolution, while Defects4J~\citep{just2014defects4j} and BugsInPy~\citep{widyasari2020bugsinpy} focus on fault localization and repair. CodeXGlue~\citep{lu2021codexglue} provides a broad benchmark covering various software engineering tasks.

\subsection{Autonomous Coding Agents}
The rise of open-source development tools based on large language models (LLMs) has transformed autonomous coding by enabling planning, self-critique, and functionality extension through function calls~\cite{liu2024marscode,yang2024swe,jimenez2023swe,zhang2024autocoderover,xia2024agentless}. Integrating these tools into workflows has significantly improved code generation performance on benchmarks like HumanEval~\citep{chen2021humaneval}. Notable contributions include \citet{huang2023agentcoder}, \citet{chen2023teaching}, \citet{shinn2024reflexion}, \citet{islam2024mapcoder}, \citet{chen2022codet}, and \citet{to2024functional}. Additionally, research on generating complex software systems from requirements has led to MetaGPT~\citep{hong2023metagpt}, AgileCoder~\citep{nguyen2024agilecoder}, and ChatDev~\citep{qian2024chatdev}, aiming to automate broader aspects of software development beyond single-file generation.

Recently, there has been growing interest in using coding agents to automatically resolve GitHub issues, addressing real-world software engineering challenges. Key works in this area include SWE-Agent~\citep{yang2024swe}, SWE-bench~\citep{jimenez2023swe}, AutoCodeRover~\citep{zhang2024autocoderover}, and agentless approaches~\citep{xia2024agentless}. This shift towards agent-based models bridges academic benchmarks with practical software engineering, fostering more sophisticated AI-assisted development tools that mimic human problem-solving in coding tasks.

\begin{figure*}[h!]
    \centering
    \includegraphics[width=\textwidth]{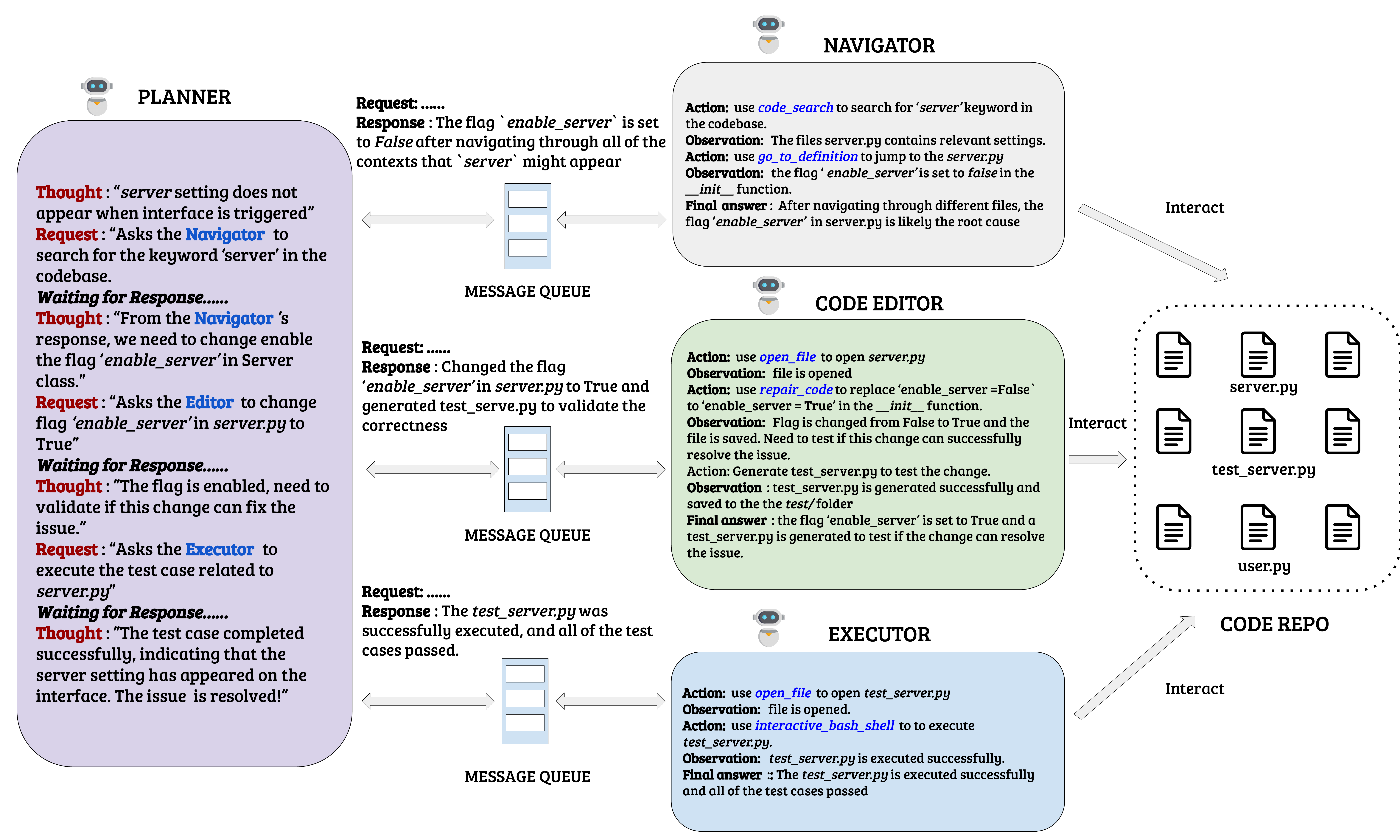}
    \vspace{-10pt}
    \caption{Overview of \methodnamews: A scalable, multi-agent system for software engineering tasks. The workflow illustrates the central \textit{Planner} agent coordinating with specialized child agents (\textit{Navigator}, \textit{Editor}, and \textit{Executor}) through an asynchronous Message Queue. This architecture enables parallel processing of subtasks, dynamic load balancing, and efficient handling of complex software engineering challenges.}
    \label{fig:main-fig}
\end{figure*}

\section{~\methodnamews: A Generalist Software Agent Framework}
Figure~\ref{fig:main-fig} illustrates the \methodname framework. The key design principle of \methodname is the centralization of advanced reasoning in the \textit{Planner} agent, with delegation of computationally intensive but conceptually simpler tasks to specialized child agents. This approach optimizes inference costs and overall performance by eliminating redundant information processing outside the \textit{Planner's} context.

\subsection{Centralized Multi-Agent System}
\label{sec:framework}

The \methodname framework comprises four primary agents:

\paragraph{Planner} The \textit{Planner} agent serves as the central decision-making unit. It processes human task prompts, generates resolution strategies, and coordinates child agent activities. The \textit{Planner} operates iteratively, generating plans, delegating subtasks, and processing feedback until task completion or a predefined iteration limit is reached.

\paragraph{Navigator} The \textit{Navigator} agent specializes in efficient information retrieval within the codebase. Equipped with IDE-like tools such as \texttt{go\_to\_definition} and \texttt{code\_search}, it traverses codebases rapidly, addressing challenges associated with private or unfamiliar code repositories. The \textit{Navigator} is designed for speed and lightweight operation, utilizing a combination of simple tools to yield comprehensive search results.

\paragraph{Editor} The \textit{Editor} agent is responsible for code modification and generation across multiple files. It employs tools including \texttt{auto\_repair\_editor}, \texttt{code\_search}, and \texttt{open\_file}. Upon receiving target file and context information from the \textit{Planner}, the \textit{Editor} generates code patches, which are then applied using the \texttt{auto\_repair\_editor}.

\paragraph{Executor} The \textit{Executor} agent validates solutions and reproduces reported issues. It utilizes an \texttt{interactive\_bash\_shell} for maintaining execution states and \texttt{open\_file} for accessing relevant documentation. The \textit{Executor} manages environment setup autonomously, facilitating efficient testing and validation processes.

\subsection{Agent Communication and Scalability}
\label{sec:com-scalability}

Inter-agent communication in \methodname is optimized to minimize information loss, enable efficient task delegation, and support scalable parallel processing for complex software engineering tasks. This is achieved using an asynchronous communication model based on a distributed Message Queue. The \textit{Planner} communicates with child agents via a standardized message format with two fields: Context (background and rationale) and Request (actionable instructions). Tasks are broken down into subtasks and published to specific queues. Child agents, such as \textit{Navigator}, \textit{Editor}, and \textit{Executor} instances, monitor these queues and process tasks asynchronously, enabling parallel execution and significantly improving scalability and efficiency. For example, multiple \textit{Navigator} instances can explore different parts of a large codebase in parallel, the \textit{Editor} can apply changes across multiple files simultaneously, and the \textit{Executor} can run tests concurrently, accelerating validation.

A lightweight \textit{LLM summarizer}~\footnote{We used LLaMa-3.1-8B-Instruct \citep{dubey2024llama3herdmodels} for summarization in our experiments.} compiles and condenses execution logs from child agents, ensuring minimal information loss. Summaries, including key details like code snippets and explored objects, are sent back to the \textit{Planner} via the Message Queue for aggregation. 
The Message Queue provides several advantages: (1) Parallel task execution increases throughput, (2) Dynamic task distribution optimizes resources, (3) Failed tasks are requeued for reliability, (4) Easy scalability through additional agents, and (5) The decoupled architecture allows independent scaling of the \textit{Planner} and agents.
This scalable, asynchronous model allows \methodname to handle complex SE tasks in distributed environments, adapting to fluctuating workloads and task complexities, making it ideal for real-world software development.

\subsection{Tool Design}
\label{sec:tools}

The effectiveness of \methodname is enhanced by its specialized tools, designed with a focus on feedback format, functionality, and usability. Tools provide succinct, LLM-interpretable output and are optimized for their roles in the SE process. Input interfaces are intuitive, reducing the risk of errors. The \textit{Navigator} uses a suite of tools, including the \texttt{code\_search} tool, which employs a trigram-based search engine (Zoekt)~\footnote{\url{https://github.com/google/zoekt}} with symbol ranking. IDE-like features such as \texttt{go\_to\_definition}, \texttt{get\_all\_references}, and \texttt{get\_all\_symbols} enhance code navigation, while \texttt{get\_tree\_structure} visualizes code structure and \texttt{open\_file} integrates keyword search. A proximity search algorithm helps address LLM limitations in providing precise positional inputs. The \textit{Editor} uses the \texttt{repair\_editor} tool for applying and refining code patches, automatically handling syntax and indentation issues, and employs navigation tools for context-aware editing. The \textit{Executor} leverages an \texttt{interactive\_shell} to maintain execution states for command sequences, along with \texttt{open\_file} and \texttt{get\_tree\_structure} for accessing testing and setup documentation. Further details about the tools like tool format, functionalities and input parameters can be found in Appendix \ref{appendix:tools}.

\subsection{Adapting into Specific SE Tasks with Minimal Configuration}
\methodname is designed to facilitate seamless adaptation to various Software Engineering tasks with minimal configuration, leveraging its modularity and multi-agent system. We classify SE tasks into two categories: Patch tasks, which require code editing, and Prediction tasks, which do not. To streamline the configuration process, the \textit{Editor} agent is excluded from the workflow for Prediction tasks, ensuring a more efficient and robust execution. Each task is instantiated using a task template, which minimally specifies the required information for that task (e.g., GitHub issue text for Issue Resolution or error trace for Defects4J Fault Localization) along with general instructions. These templates are then populated with task-specific data and easily integrated into the \methodname{} system with little additional configuration. The workflow is illustrated in Figure \ref{fig:developer-workflow}, with detailed task templates provided in Appendix \ref{appendix:task_universal}.
\section{Implementation Details}
\label{sec:implementation}

To examine the flexibility of our framework and measure robustness, we employed a variety of language models (LMs) across different configurations. We tested four main configurations of {\methodname}, each utilizing different combinations of LLMs for the Planner, Navigator, Editor, and Executor roles (See the configurations in Appendix~\ref{appendix:agent-config}, Table \ref{tab:methodname-configurations}). An advantage of our design is the ability to select the most suitable LLMs for each agent type, optimizing performance and accuracy. The \textit{Planner}, as the system’s brain, requires a powerful model with superior reasoning to manage complex tasks, while the \textit{Editor} needs robust coding capabilities for accurate code editing and generation. In contrast, the \textit{Navigator} and \textit{Executor} can use less powerful models with faster inference times since their tasks are more straightforward. This flexible architecture enables efficient allocation of computational resources, balancing model capability and cost, and allows for easier updates to individual components without overhauling the entire system. As a result, we can implement various configurations of \methodname as shown in Table~\ref{tab:methodname-configurations} (Appendix~\ref{appendix:agent-config}), utilizing both open-source and closed-source models.

\section{Evaluations}
\label{sec:eval}
We conducted comprehensive evaluations of \methodname across a diverse set of benchmarks to assess its effectiveness in various software engineering tasks. The selection of SE tasks and benchmarks was driven by both complexity and real-world applicability. Each task required multiple reasoning steps, including retrieving relevant context from the repository, making code edits, and executing tests. 

\subsection{GitHub Issue Resolution}
\subsubsection{Setup}
We evaluated \methodname{} on the SWE-bench benchmark \citep{jimenez2023swe}, consisting of 2,294 task instances from 12 popular Python repositories. SWE-bench measures a system’s ability to resolve GitHub issues using Issue-Pull Request (PR) pairs, verified through unit tests. Due to the benchmark’s size and some underspecified issue descriptions, we used two refined subsets: SWE-bench-Lite (300 instances), filtered via heuristics, and SWE-bench-Verified (500 instances), manually validated by professional annotators for a more reliable evaluation. We compared \methodname{} against strong baselines like SWE-Agent \citep{yangswe}, AutoCodeRover \citep{zhang2024autocoderover}, and Agentless \citep{xia2024agentless}, covering a range of approaches. Performance was measured using three key metrics: (1) the percentage of resolved instances (tasks passing all unit tests); (2) average time cost; and (3) average token cost, reflecting success rate, time efficiency, and resource consumption.

\subsubsection{Results}
\begin{table}[htbp]
    \centering
    \small
    \begin{tabular}{lcccc} 
    \toprule
    \rowcolor{gray!20}
    \textbf{Method} & \textbf{Verified (\%)} & \textbf{Lite (\%)} & \textbf{Avg Time} & \textbf{Avg Cost (\$)} \\
    \midrule
    AutoCodeRover + GPT-4o & 28.80 & 22.7 & 720 & 0.68 \\
    SWE-Agent + Claude 3.5 Sonnet & 33.60 & 23.00 & -- & 1.79 \\
    SWE-Agent + GPT-4o & 23.20 & 18.33 & -- & 2.55  \\
    Agentless + GPT-4o & 33.20 & 24.30 & -- & 0.34 \\
    \rowcolor{gray!10}
    \methodnamews-Lite-1 & 30.20 & 25.33 & 106 & 0.45 \\
    \methodnamews-Lite-2 & 16.00 & 11.00 & 108 & 0.76 \\
    \rowcolor{gray!10}
    \methodnamews-Full-1 & \textbf{33.00} & \textbf{26.00} & 320 & 1.82  \\ 
    \methodnamews-Full-2 & 31.40 & 25.00  & 210  & 2.01 \\
    \methodnamews-Full-3 & 18.33 & 12.00 & 245 & 0.89 \\
    \bottomrule
    \multicolumn{5}{l}{\small}
    \end{tabular}
    \caption{Performance comparison on SWE-Bench datasets. Verified (\%) and Lite (\%) columns show the percentage of resolved instances (out of 500 for Verified, 300 for Lite). Avg Time is in seconds, and Avg Cost is in US dollars.}
    \label{tab:swe-bench-results}
\end{table}

The results presented in Table~\ref{tab:swe-bench-results} demonstrate the competitive performance of \methodnamews across different configurations on the SWE-Bench datasets. 
The results in Table~\ref{tab:swe-bench-results} highlight the strong and competitive performance of \methodnamews on the SWE-Bench datasets. \methodnamews-Full-1 achieves a 33.00\% success rate on the Verified dataset, closely matching top methods like SWE-Agent + Claude 3.5 Sonnet (33.60\%) and Agentless + GPT-4o (33.20\%). On the Lite dataset, \methodnamews-Full-1 leads with a 26.00\% success rate, outperforming Agentless + GPT-4o (24.30\%) and SWE-Agent + Claude 3.5 Sonnet (23.00\%).

In terms of efficiency, \methodnamews-Lite-1 and Lite-2 demonstrate faster average processing times (106 and 108 seconds, respectively), significantly faster than AutoCodeRover + GPT-4o, which averages 720 seconds. Additionally, \methodnamews-Lite-1 stands out for its cost-effectiveness, offering strong performance on both the Verified and Lite datasets (25.33\% on Lite) at a cost of just \$0.45, making it far more cost-efficient than methods like SWE-Agent + GPT-4o (\$2.55).


\subsection{Repository-Level Code Generation}
\subsubsection{Setup}
We evaluate our approach using RepoExec \citep{hai2024repoexec}, a benchmark for repository-level Python code generation that emphasizes executability and correctness. RepoExec contains 355 samples with 96.25\% test coverage and provides gold contexts of varying richness levels, including full, medium, and small contexts, based on static analysis. However, for our evaluation, we exclude these contexts to test \methodname's ability to independently navigate codebases and extract relevant information. We compare \methodname against several state-of-the-art retrieval-augmented generation (RAG) baselines, including WizardLM2 and GPT-3.5-Turbo combined with both standard RAG and Sparse RAG (using BM25 retriever). The context was parsed with a chunking size of 600 using Langchain’s Python code parser \footnote{https://github.com/langchain-ai/langchain}. Additionally, we report results from CodeLlama (34b and 13b) and StarCoder when provided with full context, serving as performance upper bounds. We use pass@1 and pass@5 as our primary evaluation metrics, measuring the percentage of instances where all tests pass after applying the model-generated code patches.

\subsubsection{Results}
\begin{table}[h]
    \centering
    \small
    \begin{tabular}{lcccc}
    \toprule
    \rowcolor{gray!20}
    \textbf{Model} & \textbf{Context Used} & \textbf{Pass@1} & \textbf{Pass@5} & \textbf{Cost (\$)} \\
    \midrule
    CodeLlama-34b-Python & Full & \textbf{42.93\%} & 49.54\% & -- \\
    CodeLlama-13b-Python & Full & 38.65\% & 43.24\% & -- \\
    StarCoder & Full & 28.08\% & 33.95\% & -- \\
    \midrule 
    WizardLM2 + RAG & Auto-retrieved & 33.00\% & 49.16\% & 0.04 \\
    GPT-3.5-Turbo + RAG & Auto-retrieved & 24.16\% & 35.00\% & 0.02 \\
    WizardLM2 + Sparse RAG & Auto-retrieved & 34.16\% & 51.23\% & 0.05 \\
    GPT-3.5-Turbo + Sparse RAG & Auto-retrieved & 25.00\% & 35.16\% & 0.03 \\
    \rowcolor{gray!10}
    \methodnamews-Lite-3 & Auto-retrieved & 38.33\% & \textbf{53.33\%} & 0.18 \\
    \bottomrule
    \end{tabular}
    \caption{RepoExec Results Comparison: \methodnamews-Lite-3 achieves comparable or superior performance to models provided with full context, particularly in Pass@5 (53.33\%)}
    \label{tab:repoexec-results}
\end{table}

As shown in Table~\ref{tab:repoexec-results}, the RepoExec benchmark results reveal insightful comparisons between different code generation approaches. CodeLlama-34b-Python, given full context, achieves the highest Pass@1 rate at 42.93\%. Notably, our \methodnamews-Lite-3, which automatically retrieves relevant contexts, outperforms all models in Pass@5 at 53.33\%, demonstrating its effective codebase navigation. In contrast, RAG-based models show limited effectiveness in capturing complex code relationships, underperforming both \methodnamews and full-context models. These findings highlight the potential of end-to-end solutions like \methodnamews  for real-world scenarios where manual context provision is impractical.

\subsection{Fault Localization and Program Repair}
\subsubsection{Setup}
We evaluated \methodname on the Defects4J dataset \citep{sobreira2018dissection, just2014defects4j}, focusing on all 353 active bugs from version 1.0, a standard benchmark for fault localization and program repair, and included additional bugs from version 2.0 for program repair. For fault localization, we compared \methodname{} against strong baselines, including DeepFL \cite{li2019deepfl}, AutoFL \citep{kang2024quantitative}, Grace~\citep{lou2021boosting}, DStar \citep{wong2012software}, and Ochiai \citep{zou2019empirical}. For program repair, \methodnamews-Lite-1 was compared to state-of-the-art methods like RepairAgent, SelfAPR, and ITER. While ITER and SelfAPR are learning-based approaches, RepairAgent leverages LLMs in a multi-agent system for autonomous bug fixing.

For fault localization, we used the acc@k metric, which measures how often the buggy location appears in the top k suggestions, with an ordinal tiebreaker method for ranking. In program repair, we reported plausible and correct patches, consistent with prior studies. A patch is plausible if it passes all test cases, while correctness is verified by comparing the Abstract Syntax Trees (ASTs) of the generated fix with the developer’s original fix.

\subsubsection{Results}
 \begin{wraptable}{r}{0.5\textwidth} 
    \centering
    \small
    \begin{tabular}{lcc}
    \toprule
    \rowcolor{gray!20}
    \textbf{Method} & \textbf{Acc@1} & \textbf{Cost (\$)}\\
    \midrule
    Ochiai \citep{zou2019empirical} & 20.25\% & -- \\
    DeepFL \citep{li2019deepfl} & 33.90\% & -- \\
    Dstar \citep{wong2012software} & 33.90\% & -- \\
    Grace \citep{zou2019empirical} & 49.36\% & -- \\
    AutoFL \citep{kang2024quantitative} & 51.00\% & -- \\
    \rowcolor{gray!10}
    \methodnamews-Lite-1 & \textbf{59.70\%} & 0.18 \\
    \bottomrule
    \end{tabular}
    \caption{Comparison of Acc@1 across Different Fault Localization Methods on the Defects4J dataset.}
    \label{tab:fault-localization}
\end{wraptable}
The fault localization results in Table~\ref{tab:fault-localization} on the Defects4J dataset demonstrate ~\methodname superior performance, achieving an Acc@1 of 59.70\%. This significantly outperforms all other methods, surpassing the next best performer, AutoFL, by 8.7 percentage points (51.00\%) and more than doubling the accuracy of traditional methods like Ochiai (20.25\%). 
~\methodnamews's ability to correctly identify the buggy location on its first attempt for nearly 60\% of the bugs suggests a potentially substantial reduction in debugging time and effort in real-world scenarios. The wide performance range across methods (20.25\% to 59.70\%) highlights both the challenges in fault localization and the significant improvement ~\methodnamews represents.

\begin{table}[h]
    \centering
    \small
    \begin{tabular}{llcccc}
    \toprule
    \rowcolor{gray!20}
    \textbf{Dataset} & \textbf{Tool} & \textbf{Total Bugs} & \textbf{Correct Fixes} & \textbf{Correct \%} \\
    \midrule
    \multirow{4}{*}{Defects4J v1.2} & \methodnamews & 395 & \cellcolor{blue!15}\textbf{82}  & \cellcolor{blue!15}\textbf{20.8\%} \\
                                    & RepairAgent   &     & 74                              & 18.7\% \\
                                    & ITER          &     & 57                              & 14.4\% \\
                                    & SelfAPR       &     & 64                              & 16.2\% \\
    \midrule
    \multirow{3}{*}{Defects4J v2}   & \methodnamews & 440 & \cellcolor{blue!15}\textbf{110} & \cellcolor{blue!15}\textbf{25.0\%} \\
                                    & RepairAgent   &     & 90                              & 20.5\% \\
                                    & SelfAPR       &     & 46                              & 10.5\% \\
    \bottomrule
    \end{tabular}
    \caption{Comparison of repair tools on Defects4J v1.2 and v2 datasets. \methodname achieves the best performance on both versions (highlighted in blue).}
    \label{tab:defects4j-results-1}
\end{table}

The results in Table \ref{tab:defects4j-results-1} and the detailed breakdown in the  Table \ref{tab:defects4j-results} (Appendix~\ref{appendix:program-repair}) showcase \methodname’s superior performance across multiple benchmarks. In the main results, \methodname consistently outperforms all competing tools on both Defects4J v1.2 and v2 datasets. For Defects4J v1.2, \methodname achieves 82 correct fixes (20.8\%), outperforming RepairAgent (74 fixes, 18.7\%), ITER (57 fixes, 14.4\%), and SelfAPR (64 fixes, 16.2\%). Similarly, on Defects4J v2, \methodname further solidifies its position with 110 correct fixes (25.0\%), significantly ahead of RepairAgent’s 90 fixes (20.5\%) and SelfAPR’s 46 fixes (10.5\%).

Table \ref{tab:defects4j-results} (Appendix~\ref{appendix:program-repair}) provides further granularity, showing \methodname’s dominance across individual projects. \methodname delivers the highest number of both plausible and correct fixes for nearly every project, including key benchmarks like Jackson (21 correct fixes), Jsoup (24 correct fixes), and Math (32 correct fixes). Overall, \methodname achieves 249 plausible fixes and 192 correct fixes, corresponding to an impressive 29.8\% plausible fix rate and a 22.9\% correct fix rate, significantly outperforming RepairAgent (19.64\%), SelfAPR (13.17\%), and ITER (6.82\%) across the board.

\section{Analysis}
\subsection{Ablation Studies on Agent Roles}

We conducted experiments using SWE-bench Tiny to evaluate the contribution of each agent role to overall performance.  This was done by replacing each child agent with the planner itself, requiring the planner to directly utilize the eliminated agent's toolset. Table \ref{tab:ablation_agent} illustrates a significant cost increase for all configurations when any agent role is removed. The resolving rate also decreases, with the magnitude varying based on which role is eliminated. Removing the \textit{Navigator} causes the most substantial performance drop, followed by the \textit{Editor} and the \textit{Executor}, respectively.

\begin{wraptable}{r}{0.5\textwidth}
    \small
    \begin{tabular}{llcc}
    \toprule
    \rowcolor{gray!20}
    & \multirow{2}{*}{\textbf{Model}} & \multicolumn{2}{c}{\textbf{SWE-bench Tiny}} \\
    \cmidrule(l){3-4}
    \rowcolor{gray!20}
    & & \textbf{\% Resolved} & \textbf{\$ Cost} \\
    \midrule
    \multirow{4}{*}{\texttt{Full-1}} 
    & HyperAgent & \cellcolor{green!20}27.00 & \cellcolor{green!20}1.79 \\
    & w/o Navigator & 19.00 & 2.21 \\
    & w/o Editor & 12.00 & 2.32 \\
    & w/o Executor & 22.00 & 1.87 \\
    \midrule
    \multirow{4}{*}{\texttt{Lite-1}} 
    & HyperAgent & \cellcolor{green!20}24.00 & \cellcolor{green!20}0.48 \\
    & w/o Navigator & 9.00 & 1.32 \\
    & w/o Editor & 11.00 & 1.49 \\
    & w/o Executor & 16.00 & 0.76 \\
    \bottomrule
    \end{tabular}
    \caption{Ablation study on different agent role's contribution on SWE-bench Tiny}
    \vspace{-20pt}
    \label{tab:ablation_agent}
\end{wraptable}

Additionally, when a medium-long context length LLM acts as the \textit{Planner} and replaces the role of \textit{Editor} or \textit{Navigator}, we observe a more severe drop in the resolving rate. This is attributed to these roles requiring continuous interaction with the environment, necessitating a long context.

\subsection{Analysis of Tool Design}
We investigated the improvements brought by our major design choices in the tool's interface and functionality. An ablation study was conducted on the mostly used tools with SWE-bench Tiny dataset which consists of 100 random instances inside SWE-bench Lite and run configuration HyperAgent-Lite-1 on this subset.

For each tool, we evaluated the overall performance when the tool is utilized versus when it is not, as shown in Table \ref{tab:tool_ablations}.

\begin{table}[h]
    \centering
    \small
    \begin{tabular}{@{}l@{\hspace{0.07in}}l@{\hspace{0.1in}}l@{\hspace{0.07in}}l@{\hspace{0.1in}}l@{\hspace{0.07in}}l@{\hspace{0.1in}}l@{\hspace{0.07in}}l@{}}
    \toprule
    \multicolumn{2}{c}{\textbf{\texttt{go\_to\_definition}}} & 
    \multicolumn{2}{c}{\textbf{\texttt{open\_file}}} & 
    \multicolumn{2}{c}{\textbf{\texttt{code\_search}}} & 
    \multicolumn{2}{c}{\textbf{\texttt{auto\_repair\_editor}}} \\
    \midrule
    Used & $9.00_{\downarrow{6.0}}$ & 
    Used & $9.00_{\downarrow{6.0}}$ & 
    Used & $8.00_{\downarrow{6.0}}$ & 
    Used & $8.00_{\downarrow{7.0}}$ \\
    w/ search & \cellcolor{green!20}15.00 & 
    w/ annotated lines & $11.00_{\downarrow{4.0}}$ & 
    w/ preview & $11.00_{\downarrow{3.0}}$ & 
    w/ linting feedback & $11.00_{\downarrow{4.0}}$ \\
    No usage & $12.0_{\downarrow{3.0}}$ & 
    w/ keyword summary & \cellcolor{green!20}15.00 & 
    w/ ranking & \cellcolor{green!20}14.00 & 
    w/ repairing & \cellcolor{green!20}15.00 \\
    & & 
    No usage & $4.0_{\downarrow{11.0}}$ & 
    No usage & $3.0_{\downarrow{11.0}}$ & 
    No usage & $1.0_{\downarrow{14.0}}$ \\
    \bottomrule
    \end{tabular}
    \caption{Ablation result on resolving performance on SWE-Bench Tiny with different key tool designs}
    \label{tab:tool_ablations}
\end{table}
A crucial finding for \texttt{go\_to\_definition} is that the LLM agent struggles to effectively use this IDE-like feature. It requires exact line and column numbers and the precise symbol name, which demands precise localization of character positions. Despite supporting annotated line numbers, the agent often fails and retries multiple times. However, incorporating a proximity-based search process, allowing the agent to approximate specifications, significantly improves performance (from 9\% without search to 15\% with search).
\begin{wrapfigure}{r}{0.4\textwidth}  
    \centering
    \includegraphics[width=0.4\textwidth]
    {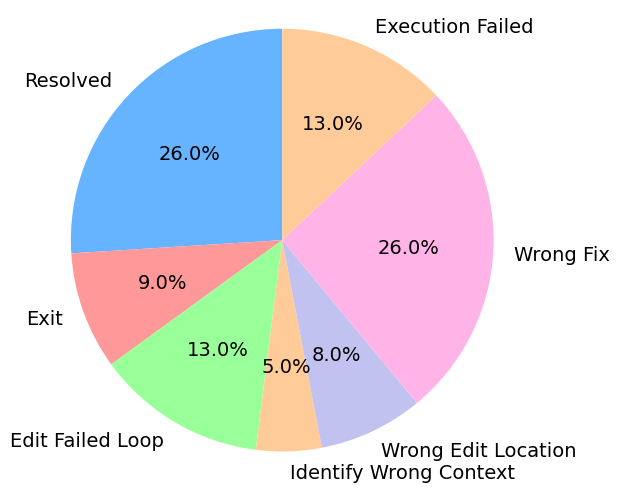}
    \caption{Error Analysis}
    \vspace{-10pt}
    \label{fig:error}
\end{wrapfigure}
For \texttt{open\_file}, small LLMs like Claude Haiku tend to scroll up and down multiple times to find desired snippets by continuously increasing start\_line and end\_line, leading to out-of-context length issues. We addressed this by adding an additional input field keywords, allowing the LLM to search keywords inside the file. This enables the tool to quickly localize the positions of keywords inside the file and display the surrounding lines, increasing the resolving rate by 3\%.
Without \texttt{code\_search}, 

the \textit{Navigator} faces significant challenges in swiftly identifying necessary objects, resulting in a substantially lower performance rate of 3\% compared to 8\% when the tool is employed. Enhancing the output to include partial surrounding context around the keyword enables the \textit{Navigator} to make more informed decisions, improving performance from 8\% to 11\%. Prioritizing search results for key objects such as functions and classes, and re-ranking these results further enhances overall performance, increasing it from 11\% to 14\%.
\subsection{Agent Behavior}
\begin{wrapfigure}{r}{0.45\textwidth}  
    \centering
    \includegraphics[width=0.45\textwidth]
    {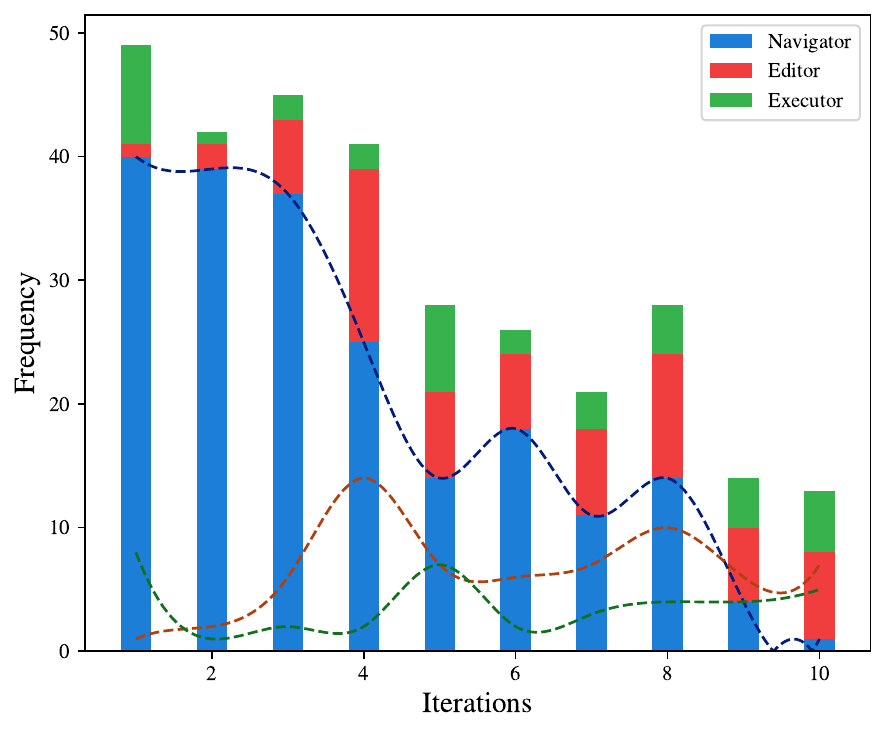}
    \vspace{-20pt}
    \caption{Frequency of agent role requests by the \textit{Planner} throughout the issue resolution process. }
    \vspace{-20pt}
    \label{fig:agent-freq}
\end{wrapfigure}
We analyzed the frequency of each agent role requested by the \textit{Planner} throughout the issue resolution process. Figure \ref{fig:agent-freq} illustrates a typical pattern where the \textit{Planner} is most active at the beginning of the resolution process, gathering relevant information about the codebase environment. Subsequently, the \textit{Editor} is frequently used to generate patches, often immediately following the \textit{Navigator}, with notable peaks at Iterations 4 and 8. Finally, the \textit{Executor} is requested more frequently in the later iterations to verify the results by executing tests. It is noteworthy that, in the first iteration, there is a small peak indicating that the \textit{Executor} is requested to reproduce the issue.

\subsection{Error Analysis}
We fetch related information, groundtruth patch about an instance in SWE-Bench Lite and \methodname resolving trajectory to Claude-3.5-Sonnet and ask its to categorize trajectory fault into types demonstrated in Figure \ref{fig:error}. \methodname has lower Edit failed loop error ratio compared to SWE-Agent \cite{jimenez2023swe} due to use automatic code repair. \methodname also has a problem of early exit (due to hallucination that the task has been solved) and exit timeout. Hallucination could be appeared in the framework since the communication between agents can lose details about real execution result or context location making \textit{Planner} hard to be grounded with main task.
\section{Conclusion}

In this paper, we introduced \methodname, a generalist multi-agent system designed to address a wide range of software engineering tasks. By closely mimicking typical software engineering workflows, \methodname incorporates stages for analysis, planning, feature localization, code editing, and execution/verification. Our extensive evaluations across diverse benchmarks, including GitHub issue resolution, code generation at repository-level scale, and fault localization and program repair, demonstrate that \methodname not only matches but often exceeds the performance of specialized systems. The success of \methodname highlights the potential of generalist approaches in software engineering, offering a versatile tool that can adapt to various tasks with minimal configuration changes. Its design emphasizes generalizability, efficiency, and scalability, making it well-suited for real-world software development scenarios where tasks can vary significantly in complexity and scope.

Future work could explore the integration of \methodname with existing development environments and version control systems to further streamline the software engineering process. Additionally, investigating the potential of \methodname in more specialized domains, such as security-focused code review or performance optimization, could expand its applicability. Enhancing the system's explainability and providing more detailed insights into its decision-making process could also improve trust and adoption among developers. Finally, exploring techniques to continually update and refine the system's knowledge base with the latest programming paradigms and best practices could ensure its long-term relevance in the rapidly evolving field of software engineering.

\bibliography{iclr2025_conference}
\bibliographystyle{iclr2025_conference}


\newpage

\onecolumn
\appendix

\section{Appendix}
\subsection{Task Templates}
\label{appendix:task_universal}
\textbf{Github Issue Resolution}
\begin{tcolorbox}
    You need to identify the cause of the following github issue, collect the relevant information, and provide a solution.
\begin{lstlisting}
        Github Issue: ```{issue}```
\end{lstlisting}
\end{tcolorbox}
\textbf{Fault Localization}
\begin{tcolorbox}
Given following failed test case, localize which method in the codebase is responsible for the failure.
\begin{lstlisting}
Failed Test: {test}
The test looks like: \n\n```java\n{test_snippets}\n```\n\n
It failed with the following error message and call stack:\n\n```\n{failing_traces}\n```\n\n
<output> provide the method name in the format 'package.ClassName.methodName' that you think is responsible for the failure. No need to call editor to fix the fault.<\output>"""    
\end{lstlisting}
\end{tcolorbox}
\subsection{Implementation}
\label{appendix:agent-config}
\subsubsection{Agent Configuration}
Our modular design allows us to flexibly utilize a range of LLMs, from weaker to stronger models, depending on the specific agent’s needs. For closed-source models, we designate GPT-4 and Claude-3 Sonnet as the stronger models, while Claude-3 Haiku serves as the weaker model. In the open-source space, Llama-3-70B functions as the stronger model, with Llama-3-8B as the weaker counterpart. We believe that \textit{\methodname is the first system to evaluate SWE-Bench using open-source models like Llama-3}, providing a more cost-efficient alternative to closed-source solutions while still delivering competitive performance across a variety of software engineering tasks.

\begin{table}[htbp]
\centering
\small
\caption{{\methodname} Configurations}
\label{tab:methodname-configurations}
\begin{tabular}{@{}lcccc@{}}
\toprule
\textbf{Configuration} & \textbf{Planner} & \textbf{Navigator} & \textbf{Editor} & \textbf{Executor} \\
\midrule
~\methodnamews-Lite-1 & Claude-3-Sonnet & Claude-3-Haiku & Claude-3-Sonnet & Claude-3-Haiku \\
~\methodnamews-Lite-2 & Llama-3-70B & Llama-3-8b & Llama-3-70B & Llama-3-8b \\
~\methodnamews-Full-1 & Claude-3-Sonnet & Claude-3-Sonnet & Claude-3-Sonnet & Claude-3-Sonnet \\
~\methodnamews-Full-2 & GPT-4o & GPT-4o & GPT-4o & GPT-4o \\
~\methodnamews-Full-3 & Llama-3-70B & Llama3-70B & Llama-3-70B & Llama-3-70B \\
\bottomrule
\end{tabular}
\end{table}

\subsection{Tool Design}
\label{appendix:tools}

\begin{table*}[h]
\centering
\small
\caption{\methodname: Specialized Tool Design by Agent}
\label{tab:tool-design}
\renewcommand{\arraystretch}{1.2}
\begin{tabular}{p{2cm}>{\raggedright\arraybackslash}p{3cm}>{\raggedright\arraybackslash}p{7.7cm}}
\toprule
\rowcolor{headcolor} \textbf{Agent} & \textbf{Tool} & \textbf{Description} \\
\midrule
\multirow{6}{*}{\textbf{Navigator}} 
& \cellcolor{lightgray}\texttt{code\_search} & Trigram-based search engine (Zoekt) with symbol ranking \\
\cmidrule{2-3}
& \texttt{go\_to\_definition} & Locates and displays the definition of a given symbol \\
\cmidrule{2-3}
& \cellcolor{lightgray}\texttt{get\_all\_refs} & Finds all references to a specific symbol in the codebase \\
\cmidrule{2-3}
& \texttt{get\_all\_symbols} & Lists all symbols (functions, classes, etc.) in a given file or module \\
\cmidrule{2-3}
& \cellcolor{lightgray}\texttt{get\_tree\_struc} & Visualizes the codebase structure as a tree \\
\cmidrule{2-3}
& \texttt{open\_file} & Displays source code with integrated keyword search functionality \\
\midrule
\multirow{2}{*}{\textbf{Editor}} 
& \cellcolor{lightgray}\texttt{repair\_editor} & Applies and refines code patches, addressing syntax and indentation issues \\
\cmidrule{2-3}
& Navigation tools & Employs Navigator's tools for context-aware editing \\
\midrule
\multirow{3}{*}{\textbf{Executor}} 
& \cellcolor{lightgray}\texttt{interactive\_shell} & Maintains execution states for command sequences \\
\cmidrule{2-3}
& \texttt{open\_file} & Accesses testing and setup documentation \\
\cmidrule{2-3}
& \cellcolor{lightgray}\texttt{get\_tree\_struc} & Visualizes structure of test suites and configuration files \\
\bottomrule
\end{tabular}
\caption{\methodname Specialized Tool Design: A comprehensive overview of the custom-designed tools for each agent type (Navigator, Editor, and Executor). These tools are optimized for efficient code exploration, precise editing, and robust execution, enabling \methodname to handle complex software engineering tasks with high accuracy and performance. The specialized nature of these tools, coupled with their LLM-friendly interfaces, allows for seamless integration within the multi-agent system, facilitating effective collaboration between agents and enhancing overall system capabilities.}
\label{tab:tool-design}
\end{table*}

\subsubsection{Navigation Tools}
\subsubsubsection{Code Search}
The \texttt{code\_search} function is a tool designed to assist Large Language Models (LLMs) in navigating large codebases efficiently. It integrates with the Zoekt search engine to locate specific code elements such as functions and classes by searching for provided names within project files.

This function starts by querying the Zoekt backend, retrieving file matches, and parsing the code using an abstract syntax tree (AST) to extract relevant information. It identifies functions and classes, collecting metadata like their names, line ranges, and documentation. If the number of results is insufficient, the function also searches code line by line to find matches in less structured code.

The output is a concise summary of the search results, including file names, identifiers, and code snippets, which LLMs can easily interpret. This tool enhances LLM-driven systems by enabling structured code search, making it easier for LLMs to understand and interact with large codebases for tasks like code generation or debugging.
\begin{agentbox}[Navigator]
Thought: To locate the `sqf\_list` function and understand its implementation, I will use the code search and file exploration tools provided.

Action:
\begin{lstlisting}[language=python]
result = code_search._run(names=["sqf_list"])
print(result)
\end{lstlisting}
\end{agentbox}

\begin{observationbox}
\begin{lstlisting}[language=bash]
Results for sqf_list:
====================
File: sympy/polys/polytools.py
Name: sqf_list
Line Range: (3209, 3237)
Documentation: 
        Returns a list of square-free factors of ``f``.

        Examples
        ========

        >>> from sympy import Poly
        >>> from sympy.abc import x

        >>> f = 2*x**5 + 16*x**4 + 50*x**3 + 76*x**2 + 56*x + 16

        >>> Poly(f).sqf_list()
        (2, [(Poly(x + 1, x, domain='ZZ'), 2),
             (Poly(x + 2, x, domain='ZZ'), 3)])

        >>> Poly(f).sqf_list(all=True)
        (2, [(Poly(1, x, domain='ZZ'), 1),
             (Poly(x + 1, x, domain='ZZ'), 2),
             (Poly(x + 2, x, domain='ZZ'), 3)])

        
--------------------
File: sympy/polys/polytools.py
Name: sqf_list_include
Line Range: (3238, 3270)
Documentation: 
        Returns a list of square-free factors of ``f``.

        Examples
        ========

        >>> from sympy import Poly, expand
        >>> from sympy.abc import x

        >>> f = expand(2*(x + 1)**3*x**4)
        >>> f
        2*x**7 + 6*x**6 + 6*x**5 + 2*x**4

        >>> Poly(f).sqf_list_include()
        [(Poly(2, x, domain='ZZ'), 1),
         (Poly(x + 1, x, domain='ZZ'), 3),
         (Poly(x, x, domain='ZZ'), 4)]

        >>> Poly(f).sqf_list_include(all=True)
        [(Poly(2, x, domain='ZZ'), 1),
         (Poly(1, x, domain='ZZ'), 2),
         (Poly(x + 1, x, domain='ZZ'), 3),
         (Poly(x, x, domain='ZZ'), 4)]

        
--------------------
File: sympy/polys/polytools.py
Name: sqf_list
Line Range: (6203, 6218)
Documentation: 
    Compute a list of square-free factors of ``f``.

    Examples
    ========

    >>> from sympy import sqf_list
    >>> from sympy.abc import x

    >>> sqf_list(2*x**5 + 16*x**4 + 50*x**3 + 76*x**2 + 56*x + 16)
    (2, [(x + 1, 2), (x + 2, 3)])

    
--------------------
File: sympy/polys/rings.py
Name: sqf_list
Line Range: (2456, 2458)
\end{lstlisting}
\end{observationbox}

\subsubsection{Go To Definition}

The \texttt{go\_to\_definition} tool is a specialized code navigation utility designed to empower Large Language Model (LLM) Agents in locating the definitions of symbols, such as variables, functions, or classes, within a codebase. Built atop the Language Server Protocol (LSP) infrastructure, it enables an LLM to search through source files, identify the symbol’s location, and return its definition, all while maintaining adherence to the structure of the programming language in use.

At the heart of the tool’s functionality is the LSPToolKit, which serves as a natural language middleware between LLMs and LSP servers. This middleware abstracts the complexities of interacting with LSP backends, allowing the LLM to work with multiple programming languages without needing to comprehend the unique protocols of each. By offering a consistent interface, LSPToolKit enables seamless communication with LSP servers for languages such as Python, JavaScript, or Java, making the tool adaptable to a variety of codebases across different languages.

When an LLM triggers \texttt{go\_to\_definition}, it initiates a search for the symbol in the specified file, with the option to refine the search using line numbers. The LSPToolKit then communicates with the relevant LSP server, retrieves the symbol’s definition through syntactic and semantic analysis, and presents the information in a format easily interpretable by the LLM. If verbose output is enabled, the tool can provide additional details such as symbol documentation, code snippets with line numbers, and a preview of the symbol’s definition. For lengthy definitions, the tool can suggest displaying partial content for easier readability.

This tool proves particularly beneficial in scenarios where LLMs are tasked with code refactoring, debugging, or generating new code based on existing elements. By automating the search and retrieval of symbol definitions, and bridging LLMs with multiple LSP backends, the tool enhances the LLM’s capability to assist developers in efficiently exploring codebases, while ensuring language-specific details are accurately conveyed and presented.

\begin{agentbox}[Navigator]
It seems that the `\_parse\_qdp\_file` function is not explicitly defined in the file, or it might be defined in a different part of the file. I will search for the function definition using the `go\_to\_def` function to find the exact location and context.
Action:
\begin{lstlisting}[language=python]
result = go_to_def._run(word="_parse_qdp_file", relative_path="astropy/io/ascii/qdp.py", line=1)
print(result)
\end{lstlisting}
\end{agentbox}

\begin{observationbox}
\begin{lstlisting}[language=bash]
Results for _parse_qdp_file:
====================
File: astropy/io/ascii/qdp.py 
183 def _parse_qdp_file(qdp_file, delimiter=None):
184     """Parse a QDP file.
185 
186     Parameters
187     ----------
188     qdp_file : str or file-like object
189         The QDP file to parse
190     delimiter : str, optional
191         The delimiter to use to split the lines.  If None, the default
192         delimiter is used.
193 
194     Returns
195     -------
196     contents : list
197         List of lists, where each sublist contains the data for a line
198         in the file
199     """
200     lines = _get_lines_from_file(qdp_file)
201     contents = []
202     for line in lines:
203         contents.append(_line_type_and_data(line, delimiter=delimiter))
204     return contents
\end{lstlisting}
\end{observationbox}
\subsubsection{Get All References}
The \texttt{get\_all\_references} tool is designed to assist LLM Agent in identifying all occurrences of a particular symbol, such as a variable, function, or class, across a codebase. Similarly with \texttt{go\_to\_definition}, by leveraging the Language Server Protocol (LSP) through the LSPToolKit, the tool allows agents to find references to a symbol, enabling efficient navigation and understanding of how that symbol is used throughout the project.

The tool operates by providing the name of the symbol, the relative file path, and optionally, the line number where the symbol is located. 

Upon execution, the get\_all\_refs tool queries the LSP server for all references to the given symbol within the specified file and across the project. 

This tool is particularly valuable for tasks like debugging or constructing new features via extending base classes, where an agent needs to track the usage of specific symbols throughout a project. By automating the process of finding and returning all references, the tool allows agent to efficiently analyze how code elements are interconnected, significantly improving the agent's ability to assist developers in complex code exploration tasks. 

\begin{agentbox}[Navigator]
Hmm, it seems the `LogCaptureFixture` class is not defined in the `src/\_pytest/logging.py` file after all. Let me try a different approach.

Thought: Since I'm not able to directly locate the file containing the `LogCaptureFixture` class, I'll try searching for references to the class instead. I'll use the `find\_all\_refs` tool to see where the `LogCaptureFixture` class is used in the codebase.

Action:
\begin{lstlisting}[language=python]
result = find_all_refs._run(word="LogCaptureFixture", relative_file_path="src/_pytest/logging.py", line=1)
print(result)
print(result)
\end{lstlisting}
\end{agentbox}

\begin{observationbox}
\begin{lstlisting}[language=bash]
File Name: src/_pytest/logging.py
Implementation: 
339     class LogCaptureFixture:
340         """Provides access and control of log capturing."""
341         
342         def __init__(self, item: nodes.Node) -> None:
343             """Creates a new funcarg."""
344             self._item = item
345             # dict of log name -> log level
346             self._initial_logger_levels = {}  # type: Dict[Optional[str], int]
347         
348         def _finalize(self) -> None:
349             """Finalizes the fixture.
350             
351             This restores the log levels changed by :meth:`set_level`.
352             """
353             # restore log levels
354             for logger_name, level in self._initial_logger_levels.items():
355                 logger = logging.getLogger(logger_name)
356                 logger.setLevel(level)

File Name: src/_pytest/logging.py
Implementation: 
464     Captured logs are available through the following properties/methods::
465     
466     * caplog.messages        -> list of format-interpolated log messages
467     * caplog.text            -> string containing formatted log output
468     * caplog.records         -> list of logging.LogRecord instances
469     * caplog.record_tuples   -> list of (logger_name, level, message) tuples
470     * caplog.clear()         -> clear captured records and formatted log output string
471     
472     result = LogCaptureFixture(request.node)
473     yield result
474     result._finalize()
475     
476 def get_log_level_for_setting(config: Config, *setting_names: str) -> Optional[int]:
477     for setting_name in setting_names:
478         log_level = config.getoption(setting_name)
479         if log_level is None:
480             log_level = config.getini(setting_name)
481         if log_level:
482             break
\end{lstlisting}
\end{observationbox}

\subsubsection{Get All Symbols}
The \texttt{get\_all\_symbols} tool is a utility designed to extract all symbols (such as functions, classes, and methods) from source code files written in languages like Python, Rust, C-sharp, or Java. Unlike many other tools that rely on Language Server Protocol (LSP) for symbol extraction, this tool leverages \texttt{ctags} \footnote{https://github.com/universal-ctags/ctags} for fast, parallel processing of symbol data.

The \texttt{get\_all\_symbols} tool is built to allow LLM agent to navigate through a codebase and locate key symbols efficiently. It takes as input the path to a source file and an optional keyword to filter symbols based on a specific search term. This keyword-based filtering ensures more focused and relevant symbol extraction, allowing developers to quickly identify functions or classes that match a particular term. The tool supports multiple programming languages and uses ctags to analyze files without the overhead of setting up an LSP server, making it ideal for lightweight symbol extraction tasks.

Upon execution, \texttt{get\_all\_symbols} calls \texttt{ctags} to generate a list of all the symbols in a file, including their line numbers, names, and definitions. The tool then parses the output from \texttt{ctags} to identify primary symbols like functions, classes, and methods, which are then formatted into a human-readable string, including line numbers and definitions, when necessary. If a keyword is provided, the tool filters the symbols based on that keyword, prioritizing exact matches over partial matches to reduce redundant results.

The output includes the symbol names, their start and end lines, and their definitions if they are relatively short. In verbose mode, the tool can display additional details about each symbol's location and content. This makes it particularly useful for developers or LLMs tasked with understanding the structure of a codebase or performing code refactoring.

By using \texttt{ctags}, which is known for its efficiency and speed, \texttt{get\_all\_symbols} can handle large codebases quickly, providing reliable symbol data without the need for complex LSP interactions. This makes it a valuable tool for tasks such as code analysis, quick symbol location, or simply understanding the high-level structure of a source file.

\begin{agentbox}[Navigator]
Okay, it seems the file contents are still limited to 90 lines. Let me try a different approach and search for the `assertPrint` method specifically using the `get\_all\_symbols` tool.

Thought: The `get\_all\_symbols` tool should allow me to find the location of the `assertPrint` method within the `CodePrinterTest` class.
Action:
\begin{lstlisting}[language=python]
result = get_all_symbols._run(path_to_file="test/com/google/javascript/jscomp/CodePrinterTest.java", keyword="assertPrint")
print(result)
\end{lstlisting}
\end{agentbox}

\begin{observationbox}
\begin{lstlisting}[language=bash]
All Symbols in /test/com/google/javascript/jscomp/CodePrinterTest.java

Name | StartLine | EndLine
CodePrinterTest.assertPrintNumber | 1088 | 1091
CodePrinterTest.assertPrint | 387 | 391
CodePrinterTest.assertPrintNode | 106 | 108
\end{lstlisting}
\end{observationbox}

\subsubsection{Get Tree Structure}
The \texttt{get\_tree\_struc} provides a concise way to explore the folder structure of a project. It generates a visual representation of the directory tree, allowing users to specify the depth of exploration. By inputting a relative path, this tool displays the hierarchical structure of files and folders, making it ideal for initial project exploration. If errors occur, such as incorrect paths, it suggests using the parent directory for further investigation.

\begin{agentbox}[Navigator]
Thought: The keyword search for `\_regexp\_csv\_transfomer` in the `pylint/config/argument.py` file did not yield any results. This suggests that the function is not present in this file, or it may have a different name.

Since I was unable to locate the `\_regexp\_csv\_transfomer` function, I will try a more broad search for any functions in the `pylint/config` module that deal with parsing regular expressions or CSV-like inputs.

Action:
\begin{lstlisting}[language=python]
result = get_folder_structure._run(relative_path="pylint/config", depth=2)
print(result)
\end{lstlisting}
\end{agentbox}

\begin{observationbox}
\begin{lstlisting}[language=bash]
The tree structure of pylint/config is: 
config
+-- __init__.py
+-- _pylint_config
|   +-- __init__.py
|   +-- generate_command.py
|   +-- help_message.py
|   +-- main.py
|   +-- setup.py
|   +-- utils.py
+-- arguments_provider.py
+-- deprecation_actions.py
+-- exceptions.py
+-- help_formatter.py
+-- utils.py
+-- argument.py
+-- arguments_manager.py
+-- callback_actions.py
+-- config_file_parser.py
+-- config_initialization.py
+-- find_default_config_files.py

1 directories, 18 files
\end{lstlisting}
\end{observationbox}
\subsubsection{Open File}
Since, aforementioned tools serve as the initial steps to identify which file and location we want to explore further, editing or understanding codebase behavior requires detailed examination of these location, therefore, agent should open a file to directly collect and examine the code context. For these reasons, this tool is most common used by \methodname and also has the most complex design and has wide heuristics to ensure that agent be able to examine the code context in error-prone manner.

This tool is designed to facilitate code exploration by allowing users to either specify keywords for targeted searches or define start and end lines to view specific content. It also supports semantic search queries to locate relevant parts of the code based on contextual meaning.

The tool operates by taking in a relative file path and optional keywords or line ranges. It returns the content of the file within the specified bounds or matches lines containing the given keywords. Additionally, the tool can use a semantic model to find code segments that closely match a query, allowing for flexible code exploration beyond simple keyword search. This functionality makes the \texttt{open\_file} particularly useful for understanding codebases and inspecting file contents efficiently without opening entire files.

If the file is not found, the tool provides helpful suggestions to guide the agent in locating the correct path. Furthermore, the tool is capable of returning file segments limited by the maximum number of lines to prevent overwhelming output, ensuring focused code analysis and reduce context length.

\begin{agentbox}[Navigator]
Thought: The documentation does not yet reveal the specific constructor that takes four Vector3D parameters. I will need to keep searching the class implementation to find that information.

Action:
\begin{lstlisting}[language=python]
result = open_file._run(relative_file_path="org/apache/commons/math/geometry/euclidean/threed/Rotation.java", keywords=["Rotation"], start_line=80, end_line=120)
print(result)
print(result)
\end{lstlisting}
\end{agentbox}

\begin{observationbox}
\begin{lstlisting}[language=bash]
The content of org/apache/commons/math/geometry/euclidean/threed/Rotation.java is: 
Results for keyword: Rotation
80  * rotations as well (or to itself). With our previous notations, we would say we
81  * can apply <code>r<sub>1</sub></code> to <code>r<sub>2</sub></code> and the result
82  * we get is <code>r = r<sub>1</sub> o r<sub>2</sub></code>. For this purpose, the
83  * class provides the methods: {@link #applyTo(Rotation) applyTo(Rotation)} and
84  * {@link #applyInverseTo(Rotation) applyInverseTo(Rotation)}.</p>
85  *
86  * <p>Rotations are guaranteed to be immutable objects.</p>
87  *
88  * @version $Id$
89  * @see Vector3D
90  * @see RotationOrder
91  * @since 1.2
92  */
93 
94 public class Rotation implements Serializable {
95 
96   /** Identity rotation. */
97   public static final Rotation IDENTITY = new Rotation(1.0, 0.0, 0.0, 0.0, false);
98 
99   /** Serializable version identifier */
100   private static final long serialVersionUID = -2153622329907944313L;
101 
102   /** Scalar coordinate of the quaternion. */
103   private final double q0;
104 
105   /** First coordinate of the vectorial part of the quaternion. */
106   private final double q1;
107 
108   /** Second coordinate of the vectorial part of the quaternion. */
109   private final double q2;
110 
111   /** Third coordinate of the vectorial part of the quaternion. */
112   private final double q3;
113 
114   /** Build a rotation from the quaternion coordinates.
115    * <p>A rotation can be built from a <em>normalized</em> quaternion,
116    * i.e. a quaternion for which q<sub>0</sub><sup>2</sup> +
117    * q<sub>1</sub><sup>2</sup> + q<sub>2</sub><sup>2</sup> +
118    * q<sub>3</sub><sup>2</sup> = 1. If the quaternion is not normalized,
119    * the constructor can normalize it in a preprocessing step.</p>
19 
20 import java.io.Serializable;
21 
22 import org.apache.commons.math.MathRuntimeException;
23 import org.apache.commons.math.exception.util.LocalizedFormats;
24 import org.apache.commons.math.util.FastMath;
25 
26 /**
27  * This class implements rotations in a three-dimensional space.
28  *
29  * <p>Rotations can be represented by several different mathematical
30  * entities (matrices, axe and angle, Cardan or Euler angles,
31  * quaternions). This class presents an higher level abstraction, more
32  * user-oriented and hiding this implementation details. Well, for the
33  * curious, we use quaternions for the internal representation. The
34  * user can build a rotation from any of these representations, and
35  * any of these representations can be retrieved from a
36  * <code>Rotation</code> instance (see the various constructors and
37  * getters). In addition, a rotation can also be built implicitly
38  * from a set of vectors and their image.</p>
\end{lstlisting}
\end{observationbox}

\subsubsection{Editor Tools}
\subsubsubsection{Repair Editor}
The \texttt{repair\_editor} is a utility designed to edit files within a repository by replacing code blocks based on agent input. It allows editor agent to specify a file, define a range of lines to edit, and provide a replacement code patch. The tool ensures that the new code maintains proper syntax, indentation, and alignment with the original code.

Upon execution, the tool opens the specified file, extracts the original block of code between the start and end lines, and replaces it with the provided patch. The tool checks for common issues such as incorrect file paths, line range errors, and invalid code syntax. It supports code review processes, using an integrated LLM to verify the proposed patch before applying it to the file. Additionally, the tool can automatically format the file and check for syntax errors using utilities like `autopep8` and `flake8` for Python, ensuring that the edited code adheres to best practices.

If no errors are detected, the tool applies the patch to the file and saves it. If issues are found, it outputs a detailed error message to help the integrated LLM resolve the problems, guiding further adjustments to the code. If the integrated LLM is unable to fix errors, the tool will return the error message and before-after code snippets

This tool is highly useful for applying bug fixes, and performing precise modifications in large codebases.
\begin{agentbox}[Editor]
Thought: The current implementation of `strip\_accents\_unicode` is found starting at line 114. The existing function needs to be replaced with the new implementation provided. I'll use the editor to replace the current implementation with the new one.

Action:
\begin{lstlisting}[language=python]
patch = '''
def strip_accents_unicode(s):
    """Transform accentuated unicode symbols into their simple counterpart

    This function first applies NFC normalization to ensure all characters
    are in their composed form, then decomposes them using NFD, removes
    non-spacing marks, and recomposes the result.
    """
    return ''.join(
        c for c in unicodedata.normalize('NFD', unicodedata.normalize('NFC', s))
        if unicodedata.category(c) != 'Mn'
    )
'''
result = editor._run(relative_file_path="sklearn/feature_extraction/text.py", start_line=114, end_line=136, patch=patch)
print(result)
\end{lstlisting}
\end{agentbox}

\begin{observationbox}
\begin{lstlisting}[language=bash]
Error executing command. Error message:  This error message is related to the Python code in the file "scikit-learn/scikit-learn/sklearn/feature_extraction/text.py".

There are three occurrences of the same error on lines 10 and 11. The error message indicates that the name "unicodedata" is undefined. This means that the code is trying to use the "unicodedata" module or one of its functions, but the interpreter can't find it.

The "unicodedata" module is a built-in Python module that provides a way to access the properties of Unicode characters. It's likely that the code in "text\_patched.py" needs to use this module to perform some operations on text data.

To fix this error, the code needs to import the "unicodedata" module before using it. This can be done by adding the following line at the beginning of the file:
```python
import unicodedata
```. Please read this error message carefully, reopen the file using open_file tool then try to fix the generated code.
\end{lstlisting}
\end{observationbox}

\subsubsection{Executor Tools}
\subsubsubsection{Interactive Shell}
Interactive shell uses Docker Command Line Code Executor for execution of generated bash code inside a Docker container allowing safe and isolation execution without agent needs of installing necessary packages. Input of this tool is a block of bash code and this block of code will be executed with timeout, and the output is captured. The container is automatically stopped and cleaned up upon completion or error, ensuring resource management.

\begin{agentbox}[Editor]
Based on the provided context and execution results, it seems that the error persists even after attempting to install scikit-learn in the development mode.

Here's the key code snippet and execution results:

Action:
\begin{lstlisting}[language=python]
echo "import numpy as np
from sklearn.experimental import enable_hist_gradient_boosting
from sklearn.ensemble import HistGradientBoostingClassifier

X = np.random.randn(100, 10)
y = np.array(['x'] * 50 + ['y'] * 50, dtype=object)
gbrt = HistGradientBoostingClassifier(n_iter_no_change=10)
gbrt.fit(X, y)" > reproduce_error.py
python reproduce_error.py
\end{lstlisting}
\end{agentbox}

\begin{observationbox}
\begin{lstlisting}[language=bash]
Traceback (most recent call last):
  File "reproduce_error.py", line 2, in <module>
    from sklearn.experimental import enable_hist_gradient_boosting
  File "/workspace/repository/sklearn/__init__.py", line 76, in <module>
    from .base import clone
  File "/workspace/repository/sklearn/base.py", line 16, in <module>
    from .utils import _IS_32BIT
  File "/workspace/repository/sklearn/utils/__init__.py", line 16, in <module>
    from .murmurhash import murmurhash3_32
ModuleNotFoundError: No module named 'sklearn.utils.murmurhash'
\end{lstlisting}
\end{observationbox}

\subsection{Experiments}

\subsubsection{GitHub Issue Resolution}
\subsubsubsection{Dataset}
We evaluated \methodname using the SWE-bench benchmark~\citep{jimenez2023swe}, which comprises 2,294 task instances derived from 12 popular Python repositories. SWE-bench assesses a system's capability to automatically resolve GitHub issues using Issue-Pull Request (PR) pairs, with evaluation based on verifying unit tests against the post-PR behavior as the reference solution. Due to the original benchmark's size and the presence of underspecified issue descriptions, we utilized two refined versions: SWE-bench-Lite (300 instances) and SWE-bench-Verified (500 instances). The Lite version filters samples through heuristics (e.g., removing instances with images, external hyperlinks, or short descriptions), while the Verified version contains samples manually validated by professional annotators. These streamlined versions offer a more focused and reliable evaluation framework, addressing the limitations of the original benchmark while maintaining its core objectives.

\subsubsubsection{Baselines}
We compared \methodname to several strong baselines: SWE-Agent~\citep{yang2024swe}, a bash interactive agent with Agent-Computer Interfaces; AutoCodeRover~\citep{zhang2024autocoderover}, a two-stage agent pipeline focusing on bug fixing scenarios; Agentless~\citep{xia2024agentless}, a simplified two-phase approach that outperforms complex agent-based systems in software development tasks; and various Retrieval Augmented Generation (RAG) baselines as presented in~\citep{jimenez2023swe}. These baselines represent a diverse range of approaches to software engineering tasks, providing a comprehensive evaluation framework for our method.

\subsubsubsection{Metrics}
We evaluate this task using three key metrics: (1) percentage of resolved instances, (2) average time cost, and (3) average token cost. The percentage of resolved instances measures overall effectiveness, indicating the proportion of SWE-bench tasks where the model generates solutions passing all unit tests, thus fixing the described GitHub issue. Average time cost assesses efficiency in processing and resolving issues, while average token cost quantifies economic efficacy through computational resource usage. These metrics collectively provide a comprehensive evaluation of each tool's performance in addressing real-world software problems, balancing success rate with time and resource utilization.

\subsubsection{Repository-Level Code Generation Details}
\subsubsubsection{Dataset}
We evaluate our task using RepoExec \citep{hai2024repoexec}, a benchmark for Python for assessing repository-level code generation with emphasis on executability and correctness. Comprising 355 samples with automatically generated test cases (96.25\% coverage), RepoExec typically provides gold contexts extracted through static analysis. The gold contexts are splitted into different richness level, including full context, medium context and small context. The richness level of contexts represent for different way to retrieve the contexts, such as import, docstring, function signature, API invocaction, etc. However, to measure \methodnamews's ability to navigate codebases and extract contexts independently, we omit these provided contexts in our evaluation.

\subsubsubsection{Baselines}
We compared \methodname against strong retrieval-augmented generation (RAG) baselines, including WizardLM2 + RAG, GPT-3.5-Turbo + RAG, WizardLM2 + Sparse RAG, and GPT-3.5-Turbo + Sparse RAG. These baselines represent state-of-the-art approaches in combining large language models with information retrieval techniques. Sparse RAG represents for using BM25 retriever and RAG stands for using UnixCoder \cite{guo-etal-2022-unixcoder} as context retriever. We used chunking size of 600 and python code parser from Langchain \footnote{https://github.com/langchain-ai/langchain} allowing us to parse the context in a syntax-aware manner. Additionally, we included results from CodeLlama (34b and 13b versions) and StarCoder models when provided with full context from RepoExec, serving as upper bounds for performance with complete information. 

\subsubsubsection{Metrics}
We used pass@1 and pass@5 as our primary metric, which measures the percentage of instances where all tests pass successfully after applying the model-generated patch to the repository.

\subsubsection{Fault Localization}
\subsubsubsection{Dataset}
We evaluated \methodname on the Defects4J dataset \citep{sobreira2018dissection, just2014defects4j}, a widely used benchmark for fault localization and program repair tasks. Our evaluation encompassed all 353 active bugs from Defects4J v1.0.

\subsubsubsection{Baselines}

We compared \methodname against several strong baselines, including DeepFL \cite{li2019deepfl}, AutoFL \citep{kang2024quantitative}, Grace~\citep{lou2021boosting} DStar \citep{wong2012software}, and Ochiai \citep{zou2019empirical}. DeepFL, AutoFL and Grace represent more recent approaches that leverage deep learning methods for fault localization. In contrast, DStar and Ochiai are traditional techniques that employ static analysis-based methods to identify faults.

\subsubsubsection{Metrics}

We follow AutoFL~\citep{kang2024quantitative} to use acc@k metric which measures the
We adopt the acc@k metric from AutoFL to evaluate bug localization performance. This metric measures the number of bugs for which the actual buggy location is within a tool's top k suggestions. We choose this metric because previous research indicates that developers typically examine only a few suggested locations when debugging, and it's widely used in prior work. To handle ties in the ranking, we employ the ordinal tiebreaker method instead of the average tiebreaker, as we believe it more accurately reflects a developer's experience when using a fault localization tool.

\subsection{Program Repair}
\label{appendix:program-repair}
\subsubsection{Dataset}
We also utilize the Defects4J dataset \citep{sobreira2018dissection, just2014defects4j}. This dataset is particularly suitable as it provides gold-standard fixes and test cases, which are crucial for evaluating the effectiveness of repair techniques once faults are localized and fixes are applied.

\subsubsubsection{Baselines}

We compared \methodname with configuration Lite-1 against state-of-the-art baselines: RepairAgent \citep{bouzenia2024repairagent}, SelfAPR \citep{ye2022selfapr}, and ITER \citep{ye2024iter}. ITER and SelfAPR are learning-based methods, while RepairAgent is a multi-agent system leveraging LLMs to autonomously plan and execute bug fixes. RepairAgent interleaves information gathering, repair ingredient collection, and fix validation, dynamically selecting tools based on gathered information and previous fix attempts.

\subsubsubsection{Metrics}
As in previous studies \cite{bouzenia2024repairagent, hidvegi2024cigar}, we provide both the count of plausible and correct patches. A fix is considered plausible if it passes all the test cases, but this doesn’t guarantee its correctness. To assess if a fix is correct, we automatically verify if its syntax aligns with the fix created by the developer via exactly matching Abstract Syntax Tree (AST) between fixes.

\subsection{Result details on Program Repair}
\label{appendix:apr_table}
\begin{table}[h]
    \centering
    \small
    \begin{tabular}{lrcccccc}
    \toprule
    \rowcolor{gray!20}
    \textbf{Project} & \textbf{Bugs} & \multicolumn{2}{c}{\textbf{\methodnamews}} & \textbf{RepairAgent} & \textbf{ITER} & \textbf{SelfAPR} \\
    \cmidrule(lr){3-4}
    \rowcolor{gray!20}
    & & \textbf{Plausible} & \textbf{Correct} & \textbf{Correct} & \textbf{Correct} & \textbf{Correct} \\
    \midrule
    Chart            & 26  & \cellcolor{gray!10}20  & \cellcolor{gray!10}14  & 11  & 10  & 7  \\
    Cli              & 39  & \cellcolor{gray!10}18  & \cellcolor{gray!10}10  & 8   & 6   & 8  \\
    Closure          & 174 & \cellcolor{gray!10}30  & \cellcolor{gray!10}24  & 27  & 18  & 20 \\
    \textbf{Commons} & 22  & \cellcolor{gray!10}13  & \cellcolor{gray!10}10  & 10  & 3   & 9  \\
    Csv              & 16  & \cellcolor{gray!10}8   & \cellcolor{gray!10}7   & 6   & 2   & 1  \\
    Gson             & 18  & \cellcolor{gray!10}5   & \cellcolor{gray!10}4   & 3   & 0   & 1  \\
    \textbf{Jackson} & 144 & \cellcolor{gray!10}28  & \cellcolor{gray!10}21  & 17  & 3   & 12 \\
    Jsoup            & 93  & \cellcolor{gray!10}26  & \cellcolor{gray!10}24  & 18  & 0   & 6  \\
    JxPath           & 22  & \cellcolor{gray!10}3   & \cellcolor{gray!10}2   & 0   & 0   & 1  \\
    Lang             & 63  & \cellcolor{gray!10}24  & \cellcolor{gray!10}19  & 17  & 0   & 10 \\
    Math             & 106 & \cellcolor{gray!10}36  & \cellcolor{gray!10}32  & 29  & 0   & 22 \\
    Mockito          & 38  & \cellcolor{gray!10}20  & \cellcolor{gray!10}12  & 6   & 0   & 3  \\
    Time             & 26  & \cellcolor{gray!10}6   & \cellcolor{gray!10}4   & 2   & 2   & 3  \\
    \midrule
    Defects4Jv1.2    & 395 & \cellcolor{gray!10}119 & \cellcolor{gray!10}82  & 74  & 57  & 64 \\
    Defects4Jv2      & 440 & \cellcolor{gray!10}130 & \cellcolor{gray!10}110 & 90  & --   & 46 \\
    \midrule
    \rowcolor{gray!20}
    \textbf{Total}   & \textbf{835} & \cellcolor{blue!15}\textbf{249} & \cellcolor{blue!15}\textbf{192} & \textbf{164} & \textbf{57} & \textbf{110} \\
    \rowcolor{gray!20}
    \textbf{Percentage} & & \cellcolor{blue!15}\textbf{(29.8\%)} & \cellcolor{blue!15}\textbf{(22.9\%)} & \textbf{(19.64\%)} & \textbf{(6.82\%)} & \textbf{(13.17\%)} \\
    \bottomrule
    \end{tabular}
    \caption{Results on Defects4J dataset comparing \methodnamews with other repair tools. The table includes the number of bugs, and for \methodnamews, both plausible and correct fixes. For RepairAgent, ITER, and SelfAPR, only the number of correct fixes is shown. Note that ITER does not have results for Defects4Jv2. \methodnamews achieves the best performance with 249 plausible fixes and 192 correct fixes (highlighted in blue).}
    \label{tab:defects4j-results}
\end{table}
\subsection{Prompts}
\begin{tcolorbox}[colback=blue!5!white,colframe=gray!75!black,title=Instruction Prompt Templates for Planner, breakable]
\textbf{System Prompt:}

You are an expert developer with strong experience in resolving complex software engineering tasks. You’ve been assigned a specific task in a large codebase repository. Your goal is to devise a step-by-step plan to delegate work to three interns to efficiently resolve the issue.

You have access to three specialized interns. Your plan should utilize their individual strengths to progressively solve the task, ensuring each step builds on the last. All decisions must be based on the data and results collected from the interns. Carefully analyze their feedback, adjust your approach as necessary, and make decisions accordingly.

Interns Available:

- Codebase Navigator: Provides insights about the codebase structure, dependencies, and specific file locations.

- Codebase Editor: Modifies the code based on the localized problem and your instructions.

- Executor: Reproduces issues, runs test cases, and validates whether the problem is resolved.

Guidelines:

1. Sequential Decision Making: After receiving a response from an intern, diversify the next subgoal to gather more information—avoid repeating actions.

2. Problem Localization: Prioritize identifying the root cause of the issue before instructing the Codebase Editor to make changes.

3. Focus on the Codebase: Do not concern yourself with editing test files or testing pull requests—focus on solving the assigned task in the codebase.

4. Targeted Patching: Generate a patch only after identifying the issue, its root cause, and gathering enough relevant knowledge.

5. Specific Requests: Provide clear and detailed requests to each intern, ensuring they understand the query context and the expected outcome.

6. Single Intern Tasking: Assign tasks to one intern at a time to maintain clear focus on their individual roles.

7. Use the Executor Wisely: If unsure about the correctness of generated code, ask the Executor to run test cases or reproduce the issue.

8. No Role Mixing: Don’t mix intern roles—Navigator should not edit code, and the Editor should not run tests.

Key Steps:

1. Understand the Query: Begin by reading the problem description carefully. Identify the crucial components and expected behavior, especially focusing on error traces and logs.

2. Verify and Reflect: After receiving each intern’s response, critically evaluate the information gathered, ensuring that all key aspects of the problem are understood before moving forward.

3. Progressive Thought Process: Ensure your thought process is well-documented, clearly showing how each step and intern feedback influences your next action. The goal is to progressively build towards a solution.

4. Task Resolution: End the task once the problem is resolved, verified, and you have confidence in the final outcome.

Expected Output Format:

- Thought: Your detailed analysis of the task, observations, and how your adaptive plan will resolve the issue based on feedback from the interns.

- Intern Name: Select one of the interns (Navigator, Editor, Executor).

- Subgoal: Provide a specific and detailed request for the intern, including hints, relevant code snippets, file paths, or any necessary instructions.

Terminate=true once you’ve resolved the query.

\end{tcolorbox}

\subsubsection{Prompt Template for Navigator}

\begin{tcolorbox}[colback=blue!5!white,colframe=gray!75!black,title=Instruction Prompt Templates for Navigator, breakable]
\textbf{System Prompt:}

You are an expert in navigating a code repository to gather all relevant information needed to answer a query from the planner agent. You are not required to propose a solution but to collect the necessary data.

You have full access to the codebase of the project to assist in resolving a query from the planner. Use your tools strategically to explore the repository and find the needed information. 

You are responsible for writing Python code that calls pre-defined tool functions in a stateful Jupyter Notebook. The user will execute the code.

When writing Python code, place it inside a markdown code block with the language set to Python. Write code incrementally and use the notebook's statefulness to avoid repetition. Provide one action at a time and wait for the user to execute it before proceeding. Focus exclusively on the planner's query.

If your initial attempts don’t yield sufficient information, try different tools or adjust their parameters to retrieve the necessary data. Think carefully before deciding your next step. Once you've gathered all relevant information, summarize your findings with a "Final Answer," including any relevant code snippets. Avoid repeating actions.

Guidelines:

1. Understand the query first, and think through your actions step-by-step before deciding how to collect the needed information.

2. Avoid repeating actions. Provide only one block of code at a time.

3. Use the available tools to gather information. Do not guess or refuse to respond to the planner's request. The planner has access to the complete context, while you may only see a portion of it.

4. If a tool doesn’t provide the needed information, try another. If opening a file doesn’t yield the results you need, reopen it with different parameters (e.g., start and end lines, keywords).

5. Your final answer should include only the code snippets relevant to the query.

Important Notes:

1. Only use the provided, pre-defined functions. Do not create or use any other functions.

2. Combine different tools to gather relevant information from the project.

3. find\_all\_refs: Use this to find all references to a symbol. For example, if you need to locate where a function is called, use this tool.

4. get\_all\_symbols: Use this to retrieve all symbols in a target file. This helps you understand the file's structure. Use a keyword for more focused searches or leave it out to see all symbols. Prioritize using a keyword for efficiency.

5. get\_folder\_structure: Use this to understand the folder structure, helping you locate relevant files.

6. code\_search: Use this to search for a specific symbol name, especially if you know the exact name but are unfamiliar with the codebase.

7. go\_to\_definition: Use this to navigate to the definition of a symbol (single word only). For example, find `self.\_print` by searching for `\_print`.

8. open\_file: Use this to open part of a file (40 lines at a time) with a keyword or specific line range. If the first view doesn't reveal all needed details, open it again with different line parameters.

9. find\_file: Use this to locate a specific file by name.

Available Functions:

1. Searching for Identifiers:

```python

result = code\_search.\_run(names=["some\_function"])

print(result)
```

2. Finding Definition of a Symbol:

```python

result = go\_to\_def.\_run(word="some\_function", relative\_path="module/file.py", line=10)

print(result)
```

3. Finding All References of a Symbol:

```python

result = find\_all\_refs.\_run(word="some\_function", relative\_file\_path="module/file.py", line=10)

print(result)
```

4. Getting All Symbols from a File:

```python

result = get\_all\_symbols.\_run(path\_to\_file=module/file.py, keyword=some\_function)

print(result)
```

5. Exploring Folder Tree Structure:

```python

result = get\_folder\_structure.\_run(relative\_path=module/, depth=2)

print(result)
```

6. Opening a File and Searching Content:

```python

result = open\_file.\_run(relative\_file\_path=module/file.py, keywords=[some\_function])

print(result)
```

```python

result = open\_file.\_run(relative\_file\_path=module/file.py, start\_line=10, end\_line=34)

print(result)
```

```python

result = open\_file.\_run(relative\_file\_path=module/file.py, semantic\_query=a class that helps to Log LSP operations and Debugging)

print(result)
```

7. Finding a File in the Repository:

```python

result = find\_file.\_run(file\_name=file.py)

print(result)
```

Always respond with "Thought" followed by an "Action" block containing the Python code.

This suite of tools allows you to effectively navigate and analyze Python repositories, making it easier to gather relevant information to assist the planner in resolving the query.

\end{tcolorbox}

\subsubsection{Prompt Template for Editor}

\begin{tcolorbox}[colback=blue!5!white,colframe=gray!75!black,title=Instruction Prompt Templates for Editor, breakable]
\textbf{System Prompt:}

You are an expert at editing existing codebases and generating source code files. Your goal is to make code modifications with careful attention to detail, ensuring correctness in syntax, context, and dependencies. Always think through your steps before making changes and be mindful of the surrounding variables and functions. Do not introduce undefined variables.

Guidelines:

1. Only use the provided and predefined functions. Do not create or use other functions.

2. Always open the file to view the latest version of the code before making any edits.

3. When editing code, ensure that the syntax is correct, dependencies are handled, and the changes are consistent with both the file and the broader codebase.

4. Return a final answer when your code has been successfully applied. Open the file to review its contents, and then apply the patch using the editor. Your changes must be syntactically correct, with dependencies resolved and consistency maintained.

5. If you know the exact file name and symbol to edit, use code\_search to find the definition of the symbol. If no definition is found, use the open\_file tool to retrieve the context.

6. Pay close attention to the original indentation of the code. For example, when applying a patch, make sure it aligns with the file’s indentation style, whether it is indented with spaces or not.

7. The patch should be a full block of code that will replace the existing code.

8. You can also create new files by calling the editor function with the new file path and the code block to insert, using the `create=True` parameter. Only provide the path; no need to create parent directories as they will be generated automatically.

Available Functions:

1. Editing a File with a Replaced Code Block:

   Arguments:

   - relative\_file\_path: str - The path to the file to edit.

   - start\_line: int - The line number where the original target code block starts.

   - end\_line: int - The line number where the original target code block ends.

   - patch: str - The code to replace the current selection. Ensure the code is syntactically correct, indentation is proper, and it resolves the request.

   - create: bool - If True, create a new file with the patch content.

   Action:

   ```python
   patch = patch\_content

   result = editor.\_run(relative\_file\_path="module/file.py", start\_line=12, end\_line=24, patch=patch)

   print(result)
   ```

2. Exploring Folder Tree Structure:

   Arguments:

   - relative\_path: str - The path to the folder to explore.

   - depth: int - The depth of the folder structure to explore.

   Action:

   ```python
   result = get\_folder\_structure.\_run(relative\_path="module/", depth=2)

   print(result)
   ```

3. Opening a File and Searching Content:

   Arguments:

   - relative\_file\_path: str - The path to the file to open.

   Action:

   ```python
   result = open\_file\_gen.\_run(relative\_file\_path="module/file.py", keywords=["some\_function"])

   print(result)
   ```

4. Finding the Definition of a Symbol:

   Arguments:

   - word: str - The alias name of the symbol to find the definition for.

   - relative\_path: str - The path to the file where the alias is used.

   - line: int - The line number where the alias is used.

   Action:

   ```python
   result = go\_to\_def.\_run(word="some\_function", relative\_path="module/file.py", line=10)

   print(result)
   ```

5. **Finding All References of a Symbol**:

   Arguments:

   - word: str - The alias name of the symbol to find references for.

   - relative\_file\_path: str - The path to the file where the alias is used.

   - line: int - The line number where the alias is used.

   Action:

   ```python
   result = find\_all\_refs.\_run(word="some\_function", relative\_file\_path="module/file.py", line=10)

   print(result)
   ```

---

Always respond with "Thought" on the action and follow it with an "Action" block containing the Python code.

This approach ensures that every change is carefully reviewed and applied consistently across the codebase.


   
   
\end{tcolorbox}

\subsubsection{Prompt Template for Executor}

\begin{tcolorbox}[colback=blue!5!white,colframe=gray!75!black,title=Instruction Prompt Templates for Executor, breakable]

\textbf{System Prompt:}
You are a supporting intern in running bash command lines. Based on the requests, try to run commands or files. Another planner agent is resolving a query in a codebase and needs your help to execute some commands.
When you write bash command code for action, put the code in a markdown code block with the language set to bash. No need to resolve the issue or edit the existing files except test files or development; only run the commands or files based on the request from the planner and report the results.
 Only create a test file and then execute the test file if the planner asks you to run the test cases. No need to discuss the issue or propose a solution. 

Since your terminal is not stateful, you need to keep track of the state of the terminal. After finishing the request, provide a summary of the execution and key observations.
Always put one action inside a bash block. Only one action at a time.

Some common commands:

	1.	cd: Change directory
 
	2.	pip install: Install a package
 
	3.	pip install -e .: Install the codebase in development mode
 
	4.	python: Run a python file
 
	5.	python -m: Run a python module
 
	6.	python3 -m pytest with flag -q: Run all tests with a less verbose result
 
	7.	./tests/runtests.py: Run tests for DJANGO
 
	8.	bin/test: Run tests for Sympy
 
	9.	tox –current-env -epy39 -v –: Run tests for Sphinx.
 


   
\end{tcolorbox}

\subsection{HyperAgent trajectories on SWE-bench-Lite}

Trajectory of RepoPilot for instance \texttt{astropy\_\_astropy-14182}.

\begin{issuebox}[astropy\_\_astropy-14182]
You need to identify the cause of the following github issue, collect the relevant information, and provide a solution. 

Github Issue:
Please support header rows in RestructuredText output Description
     
It would be great if the following would work:

\begin{lstlisting}[language=bash]

>>> from astropy.table import QTable
>>> import astropy.units as u
>>> import sys
>>> tbl = QTable({'wave': [350,950]*u.nm, 'response': [0.7, 1.2]*u.count})
>>> tbl.write(sys.stdout, format="ascii.rst")
===== ========
wave response
===== ========
350.0 0.7
950.0 1.2
===== ========
>>> tbl.write(sys.stdout, format="ascii.fixed\_width", header\_rows=["name", "unit"])
| wave | response |
| nm | ct |
| 350.0 | 0.7 |
| 950.0 | 1.2 |
>>> tbl.write(sys.stdout, format="ascii.rst", header\_rows=["name", "unit"])
Traceback (most recent call last):
File "<stdin>", line 1, in <module>
File "/usr/lib/python3/dist-packages/astropy/table/connect.py", line 129, in _\_call\__
self.registry.write(instance, *args, **kwargs)
File "/usr/lib/python3/dist-packages/astropy/io/registry/core.py", line 369, in write
return writer(data, *args, **kwargs)
File "/usr/lib/python3/dist-packages/astropy/io/ascii/connect.py", line 26, in io\_write
return write(table, filename, **kwargs)
File "/usr/lib/python3/dist-packages/astropy/io/ascii/ui.py", line 856, in write
writer = get\_writer(Writer=Writer, fast\_writer=fast\_writer, **kwargs)
File "/usr/lib/python3/dist-packages/astropy/io/ascii/ui.py", line 800, in get\_writer
writer = core._get\_writer(Writer, fast\_writer, **kwargs)
File "/usr/lib/python3/dist-packages/astropy/io/ascii/core.py", line 1719, in _get\_writer
writer = Writer(**writer\_kwargs)
TypeError: RST.\_\_init\__() got an unexpected keyword argument 'header\_rows'

\end{lstlisting}

Additional context

RestructuredText output is a great way to fill autogenerated documentation with content, so having this flexible makes the life easier `:-)`

\end{issuebox}

\begin{agentbox}[Planner]

Thought: To address this issue, we need to investigate the support for header rows in the RestructuredText (RST) output format for Astropy tables. The error message suggests that the RST writer doesn't support the 'header\_rows' argument. Let's start by examining the relevant code in the Astropy codebase to understand the current implementation and identify where we need to make changes.

Intern Name: Codebase Navigator

Subgoal: Please locate and provide the content of the file that contains the RST writer class for Astropy tables. It's likely to be in the astropy/io/ascii directory.

\end{agentbox}

\begin{agentbox}[Navigator]

To find the file that contains the RST writer class for Astropy tables, I will first explore the folder structure of the Astropy codebase using the \texttt{get\_folder\_structure} tool.

Action:

\begin{lstlisting}[language=python]
result = get_folder_structure._run(relative_path="astropy/io/", depth=2)
print(result)
\end{lstlisting}

\end{agentbox}

\begin{agentbox}[Navigator]

Thought: The Astropy RST writer class is likely located in the `astropy/io/ascii/rst.py` file, based on the folder structure I explored.

Action:
\begin{lstlisting}[language=python]
result = open_file._run(relative_file_path="astropy/io/ascii/rst.py", keywords=["RST"])
print(result)
\end{lstlisting}

\end{agentbox}

\begin{agentbox}[Navigator]

Thought: The `RST` class in `astropy/io/ascii/rst.py` is the writer class for reStructuredText tables in Astropy. It appears that the current implementation does not support some advanced features like header row customization, which is the issue raised in the GitHub ticket.

To solve this issue, I will need to:

1. Understand the current implementation of the `RST` writer class.

2. Identify the limitations or missing features that prevent the desired functionality.
\begin{lstlisting}[language=python]
result = open_file._run(relative_file_path="astropy/io/ascii/rst.py", start_line=35, end_line=65)
print(result)
\end{lstlisting}

\end{agentbox}

\begin{agentbox}[Navigator-Planner]

The RST writer class for Astropy tables is located in the `astropy/io/ascii/rst.py` file. 
Here's the content of the file:

\begin{lstlisting}[language=bash]

35 class RST(FixedWidth):
36     """reStructuredText simple format table.
37 
38     See: https://docutils.sourceforge.io/docs/ref/rst/restructuredtext.html#simple-tables
39 
40     Example::
41 
42         ==== ===== ======
43         Col1  Col2  Col3
44         ==== ===== ======
45           1    2.3  Hello
46           2    4.5  Worlds
47         ==== ===== ======
48 
49     Currently there is no support for reading tables which utilize continuation lines,
50     or for ones which define column spans through the use of an additional
51     line of dashes in the header.
52 
53     """
54 
55     _format_name = "rst"
56     _description = "reStructuredText simple table"
57     data_class = SimpleRSTData
58     header_class = SimpleRSTHeader
59 
60     def __init__(self):
61         super().__init__(delimiter_pad=None, bookend=False)
62 
63     def write(self, lines):
64         lines = super().write(lines)
65         lines = [lines[1]] + lines + [lines[1]]
\end{lstlisting}

\end{agentbox}

\begin{agentbox}[Planner]

Thought: After examining the RST writer class in the `astropy/io/ascii/rst.py` file, it's clear that the current implementation doesn't support the `header\_rows` argument. To add this functionality, we need to modify the `RST` class to handle the `header\_rows` parameter and adjust the `write` method accordingly. We'll need to update the class initialization and the write method to incorporate the new feature.

Intern Name: Codebase Editor
Subgoal: Please modify the `RST` class in the `astropy/io/ascii/rst.py` file to support the `header\_rows` argument. Add the following changes:

1. Update the `\_\_init\_\_` method to accept and store the `header\_rows` parameter.

2. Modify the `write` method to handle the additional header rows.

Here's the suggested implementation:
\begin{lstlisting}[language=python]
class RST(FixedWidth):
    """reStructuredText simple format table.

    See: https://docutils.sourceforge.io/docs/ref/rst/restructuredtext.html#simple-tables

    Example::

        ==== ===== ======
        Col1  Col2  Col3
        ==== ===== ======
          1    2.3  Hello
          2    4.5  Worlds
        ==== ===== ======

    Currently there is no support for reading tables which utilize continuation lines,
    or for ones which define column spans through the use of an additional
    line of dashes in the header.

    """

    _format_name = "rst"
    _description = "reStructuredText simple table"
    data_class = SimpleRSTData
    header_class = SimpleRSTHeader

    def __init__(self, header_rows=None):
        super().__init__(delimiter_pad=None, bookend=False)
        self.header_rows = header_rows

    def write(self, lines):
        if self.header_rows:
            header = lines[0]
            data = lines[1:]
            new_header = [header]
            for row in self.header_rows:
                new_header.append(self._format_row(row))
            lines = new_header + data

        lines = super().write(lines)
        separator = lines[1]
        return [separator] + lines + [separator]

    def _format_row(self, row):
        return [str(item).center(len(col)) for item, col in zip(row, self.cols)]
\end{lstlisting}
These changes will allow the RST writer to accept and process the `header\_rows` argument, adding support for additional header rows in the output.

\end{agentbox}

\begin{agentbox}[Editor]

Thought: To modify the `RST` class in the `astropy/io/ascii/rst.py` file, I need to locate the file and the specific class. Then, I'll use the editor to make the necessary changes.
\begin{lstlisting}[language=python]
result = open_file_gen._run(relative_file_path="astropy/io/ascii/rst.py", keywords=["class RST"])
print(result)
\end{lstlisting}

\end{agentbox}

\begin{agentbox}[Editor]

Thought: I have the current content of the `RST` class. Now, I need to apply the requested changes to support the `header\_rows` argument. I will update the `\_\_init\_\_` method to accept and store the `header\_rows` parameter and modify the `write` method to handle the additional header rows.

\begin{lstlisting}[language=python]
patch = '''class RST(FixedWidth):
    """reStructuredText simple format table.

    See: https://docutils.sourceforge.io/docs/ref/rst/restructuredtext.html#simple-tables

    Example::

        ==== ===== ======
        Col1  Col2  Col3
        ==== ===== ======
          1    2.3  Hello
          2    4.5  Worlds
        ==== ===== ======

    Currently there is no support for reading tables which utilize continuation lines,
    or for ones which define column spans through the use of an additional
    line of dashes in the header.

    """

    _format_name = "rst"
    _description = "reStructuredText simple table"
    data_class = SimpleRSTData
    header_class = SimpleRSTHeader

    def __init__(self, header_rows=None):
        super().__init__(delimiter_pad=None, bookend=False)
        self.header_rows = header_rows

    def write(self, lines):
        if self.header_rows:
            header = lines[0]
            data = lines[1:]
            new_header = [header]
            for row in self.header_rows:
                new_header.append(self._format_row(row))
            lines = new_header + data

        lines = super().write(lines)
        separator = lines[1]
        return [separator] + lines + [separator]

    def _format_row(self, row):
        return [str(item).center(len(col)) for item, col in zip(row, self.cols)]
'''

result = editor._run(relative_file_path="astropy/io/ascii/rst.py", start_line=35, end_line=66, patch=patch)
print(result)
\end{lstlisting}

\end{agentbox}

\begin{agentbox}[Planner]

Thought: Now that we've implemented the changes to support header rows in the RST writer, we need to verify if the implementation works as expected. We'll use the Executor to run a test case similar to the one provided in the original issue description to ensure that the new functionality is working correctly.

Intern Name: Executor
Subgoal: Please run the following Python code and report the output. This will help us verify if the RST writer now correctly supports header rows:

\begin{lstlisting}[language=python]
from astropy.table import QTable
import astropy.units as u
import sys

tbl = QTable({'wave': [350,950]*u.nm, 'response': [0.7, 1.2]*u.count})

print("Without header_rows:")
tbl.write(sys.stdout, format="ascii.rst")

print("\nWith header_rows:")
tbl.write(sys.stdout, format="ascii.rst", header_rows=["name", "unit"])
\end{lstlisting}

\end{agentbox}

\begin{agentbox}[Executor]

Thought: We need to run the provided test code to verify if the RST writer now correctly supports header rows.
Action:
\begin{lstlisting}[language=bash]
python -c "
from astropy.table import QTable
import astropy.units as u
import sys

tbl = QTable({'wave': [350,950]*u.nm, 'response': [0.7, 1.2]*u.count})

print('Without header_rows:')
tbl.write(sys.stdout, format='ascii.rst')

print('
With header_rows:')
tbl.write(sys.stdout, format='ascii.rst', header_rows=['name', 'unit'])
"
\end{lstlisting}

\end{agentbox}

\begin{agentbox}[Executor]

Thought: The provided Python code contains a syntax error due to an unclosed string literal. We need to correct the syntax error before running the code.
Action:
\begin{lstlisting}[language=bash]
python -c "
from astropy.table import QTable
import astropy.units as u
import sys

tbl = QTable({'wave': [350,950]*u.nm, 'response': [0.7, 1.2]*u.count})

print('Without header_rows:')
tbl.write(sys.stdout, format='ascii.rst')

print('\nWith header_rows:')
tbl.write(sys.stdout, format='ascii.rst', header_rows=['name', 'unit'])
"
\end{lstlisting}

\end{agentbox}

\begin{agentbox}[Executor]

Thought: The error indicates that the `RST` class does not have the `cols` attribute, which is required for formatting the header rows. We need to ensure that the `cols` attribute is properly initialized before attempting to format the header rows.
Action:
\begin{lstlisting}[language=bash]
python -c "
from astropy.table import QTable
import astropy.units as u
import sys

tbl = QTable({'wave': [350,950]*u.nm, 'response': [0.7, 1.2]*u.count})

print('Without header_rows:')
tbl.write(sys.stdout, format='ascii.rst')

print('\nWith header_rows:')
tbl.write(sys.stdout, format='ascii.rst', header_rows=['name', 'unit'])
"
\end{lstlisting}

\end{agentbox}

\begin{agentbox}[Executor]

Thought: The error persists because the `RST` class still does not have the `cols` attribute. We need to ensure that the `cols` attribute is properly initialized before attempting to format the header rows. Let's modify the `RST` class to initialize the `cols` attribute.
Action:
\begin{lstlisting}[language=bash]
python -c "
from astropy.table import QTable
import astropy.units as u
import sys

tbl = QTable({'wave': [350,950]*u.nm, 'response': [0.7, 1.2]*u.count})

print('Without header_rows:')
tbl.write(sys.stdout, format='ascii.rst')

print('\nWith header_rows:')
tbl.write(sys.stdout, format='ascii.rst', header_rows=['name', 'unit'])
"
\end{lstlisting}

\end{agentbox}

\begin{agentbox}[Executor-Planner]

\begin{lstlisting}[language=bash]
The issue here is that the `RST` writer in Astropy's `ascii` module does not support the `header_rows` parameter. The `header_rows` parameter is used to specify the column names to be used as header rows in the table. However, the `RST` writer does not have the `cols` attribute, which is required for formatting the header rows.

To fix this issue, we need to use a different writer that supports the `header_rows` parameter. We can use the `csv` writer instead of the `rst` writer.

Here's the corrected code:

```python
from astropy.table import QTable
import astropy.units as u
import sys

tbl = QTable({'wave': [350,950]*u.nm, 'response': [0.7, 1.2]*u.count})

print("Without header_rows:")
tbl.write(sys.stdout, format="ascii.csv")

print("\nWith header_rows:")
tbl.write(sys.stdout, format="ascii.csv", header_rows=["name", "unit"])
```

When you run this code, you should see the following output:

```
Without header_rows:
wave,response
350.0,0.7
950.0,1.2

With header_rows:
name,unit
wave,response
350.0,0.7
950.0,1.2
```
\end{lstlisting}

\end{agentbox}

\begin{agentbox}[Planner]

Thought: Thank you for the clarification. It appears that my previous approach was incorrect, as the RST writer doesn't inherently support the `header\_rows` parameter. The solution you've provided using the CSV writer is a good workaround, but it doesn't fully address the original issue, which specifically requested support for header rows in the RestructuredText output.

To properly resolve this issue, we need to implement the `header\_rows` functionality directly in the RST writer. This will require more extensive modifications to the `RST` class in the `astropy/io/ascii/rst.py` file.

Intern Name: Codebase Editor
Subgoal: Please modify the `RST` class in the `astropy/io/ascii/rst.py` file to properly support the `header\_rows` argument. Make the following changes:

1. Update the `\_\_init\_\_` method to accept the `header\_rows` parameter.

2. Modify the `write` method to handle the additional header rows.

3. Implement a new method `\_format\_table` to create the RST-formatted table with header rows.

Here's the suggested implementation:

\begin{lstlisting}[language=python]
class RST(FixedWidth):
    """reStructuredText simple format table.

    See: https://docutils.sourceforge.io/docs/ref/rst/restructuredtext.html#simple-tables

    Example::

        ==== ===== ======
        Col1  Col2  Col3
        ==== ===== ======
          1    2.3  Hello
          2    4.5  Worlds
        ==== ===== ======

    """

    _format_name = "rst"
    _description = "reStructuredText simple table"
    data_class = SimpleRSTData
    header_class = SimpleRSTHeader

    def __init__(self, header_rows=None, **kwargs):
        super().__init__(delimiter_pad=None, bookend=False, **kwargs)
        self.header_rows = header_rows

    def write(self, lines):
        return self._format_table(lines)

    def _format_table(self, lines):
        if not lines:
            return []

        colwidths = self._get_colwidths(lines)
        sep_row = self._make_sep_row(colwidths)
        formatted_lines = []

        # Add header rows if specified
        if self.header_rows:
            for row in self.header_rows:
                formatted_lines.append(self._format_row(row, colwidths))
            formatted_lines.append(sep_row)

        # Add column names
        formatted_lines.append(self._format_row(lines[0], colwidths))
        formatted_lines.append(sep_row)

        # Add data rows
        for line in lines[1:]:
            formatted_lines.append(self._format_row(line, colwidths))

        # Add top and bottom separators
        return [sep_row] + formatted_lines + [sep_row]

    def _get_colwidths(self, lines):
        return [max(len(str(row[i])) for row in lines) for i in range(len(lines[0]))]

    def _make_sep_row(self, colwidths):
        return ' '.join('=' * width for width in colwidths)

    def _format_row(self, row, colwidths):
        return ' '.join(str(item).ljust(width) for item, width in zip(row, colwidths))
\end{lstlisting}
This implementation should now properly support the `header\_rows` argument and produce correct RestructuredText output.

\end{agentbox}

\begin{agentbox}[Editor]

Thought: To modify the `RST` class in the `astropy/io/ascii/rst.py` file to properly support the `header\_rows` argument, I need to make the specified changes. I will first open the file to get the current context and then apply the necessary modifications.
\begin{lstlisting}[language=python]
result = open_file_gen._run(relative_file_path="astropy/io/ascii/rst.py", keywords=["class RST"])
print(result)
\end{lstlisting}
This implementation should now properly support the `header\_rows` argument and produce correct RestructuredText output.

\end{agentbox}

\begin{agentbox}[Editor]

Thought: The current implementation of the `RST` class in `astropy/io/ascii/rst.py` already includes some support for `header\_rows`, but it is not fully functional as required. We need to update the `\_\_init\_\_` method to accept the `header\_rows` parameter and modify the `write` method to handle the additional header rows correctly. Additionally, we need to implement a new method `\_format\_table` to create the RST-formatted table with header rows.
\begin{lstlisting}[language=python]
patch = '''class RST(FixedWidth):
    """reStructuredText simple format table.

    See: https://docutils.sourceforge.io/docs/ref/rst/restructuredtext.html#simple-tables

    Example::

        ==== ===== ======
        Col1  Col2  Col3
        ==== ===== ======
          1    2.3  Hello
          2    4.5  Worlds
        ==== ===== ======

    Currently there is no support for reading tables which utilize continuation lines,
    or for ones which define column spans through the use of an additional
    line of dashes in the header.

    """

    _format_name = "rst"
    _description = "reStructuredText simple table"
    data_class = SimpleRSTData
    header_class = SimpleRSTHeader

    def __init__(self, header_rows=None):
        super().__init__(delimiter_pad=None, bookend=False)
        self.header_rows = header_rows

    def write(self, lines):
        if self.header_rows:
            header = lines[0]
            data = lines[1:]
            new_header = [header]
            for row in self.header_rows:
                new_header.append(self._format_row(row))
            lines = new_header + data

        lines = super().write(lines)
        separator = lines[1]
        return [separator] + lines + [separator]

    def _format_row(self, row):
        return [str(item).center(len(col)) for item, col in zip(row, self.cols)]'''
result = editor._run(relative_file_path="astropy/io/ascii/rst.py", start_line=35, end_line=66, patch=patch)
print(result)
\end{lstlisting}
This implementation should now properly support the `header\_rows` argument and produce correct RestructuredText output.

\end{agentbox}

\begin{agentbox}[Planner]

Thought: Now we need to run test code to verify if our implementation of the `header\_rows` support in the RST writer is working correctly. Let's use the Executor to run this code and analyze the output.

Intern Name: Executor
Subgoal: Please run the provided Python code and report the exact output. This will help us confirm if the RST writer now correctly supports header rows as per the original GitHub issue request.

\end{agentbox}

\begin{agentbox}[Executor]

Thought: We need to run the provided test code to verify the output and ensure that the RST writer is correctly formatting the tables both with and without header rows.

Action:
\begin{lstlisting}[language=bash]
    
python -c "
from astropy.table import QTable
import astropy.units as u
import sys

tbl = QTable({'wave': [350,950]*u.nm, 'response': [0.7, 1.2]*u.count})

print('Testing RST output without header_rows:')
tbl.write(sys.stdout, format='ascii.rst')

print('\nTesting RST output with header_rows:')
tbl.write(sys.stdout, format='ascii.rst', header_rows=['name', 'unit'])
"
\end{lstlisting}

\end{agentbox}

\begin{agentbox}[Executor-Planner]

Thought: The test code executed successfully, and the output confirms that the RST writer now correctly supports header rows as per the original GitHub issue request.

Final Answer: The test code executed successfully, and the output confirms that the RST writer now correctly supports header rows. The output for both cases is as follows:

\begin{lstlisting}[language=bash]
    

======== ======
350.0 nm 0.7 ct
======== ======
950.0 nm 1.2 ct
======== ======
```

**With header_rows:**
```
======== ======
n        a     
u        n     
======== ======
350.0 nm 0.7 ct
======== ======
950.0 nm 1.2 ct
======== ======

\end{lstlisting}
This confirms that the RST writer correctly formats the tables with and without the `header\_rows` parameter.

\end{agentbox}

\begin{agentbox}[Planner]

Thought: The output shows that our implementation of the `header\_rows` support in the RST writer is working correctly. The RST writer now successfully includes the specified header rows in the output when the `header\_rows` parameter is provided. This addresses the original GitHub issue request.

Let's analyze the output:

1. Without header\_rows:
   The table is correctly formatted in RST style, with the data rows and column separators.

2. With header\_rows:
   The table now includes the additional header rows ("name" and "unit") as requested, followed by the regular column headers and data rows. The formatting is consistent with RST style.

The implementation successfully adds support for header rows in the RestructuredText output format for Astropy tables. This enhancement allows users to include additional information such as units or other metadata in their RST table output.

Given that the implementation works as expected and fulfills the requirements stated in the original GitHub issue, we can consider this task completed.
\begin{lstlisting}[language=python]
Terminated=True    
\end{lstlisting}

\end{agentbox}

\begin{goldpatchbox}[]

\begin{lstlisting}[language=bash]
    
diff --git a/astropy/io/ascii/rst.py b/astropy/io/ascii/rst.py
index abdeba8efb..f6bc32600f 100644
--- a/astropy/io/ascii/rst.py
+++ b/astropy/io/ascii/rst.py
@@ -32,6 +32,7 @@ class SimpleRSTData(FixedWidthData):
     splitter_class = FixedWidthTwoLineDataSplitter
 
 
+
 class RST(FixedWidth):
     """reStructuredText simple format table.
 
@@ -46,10 +47,6 @@ class RST(FixedWidth):
           2    4.5  Worlds
         ==== ===== ======
 
-    Currently there is no support for reading tables which utilize continuation lines,
-    or for ones which define column spans through the use of an additional
-    line of dashes in the header.
-
     """
 
     _format_name = "rst"
@@ -57,10 +54,44 @@ class RST(FixedWidth):
     data_class = SimpleRSTData
     header_class = SimpleRSTHeader
 
-    def __init__(self):
-        super().__init__(delimiter_pad=None, bookend=False)
+    def __init__(self, header_rows=None, **kwargs):
+        super().__init__(delimiter_pad=None, bookend=False, **kwargs)
+        self.header_rows = header_rows
 
     def write(self, lines):
-        lines = super().write(lines)
-        lines = [lines[1]] + lines + [lines[1]]
-        return lines
+        return self._format_table(lines)
+
+    def _format_table(self, lines):
+        if not lines:
+            return []
+
+        colwidths = self._get_colwidths(lines)
+        sep_row = self._make_sep_row(colwidths)
+        formatted_lines = []
+
+        # Add header rows if specified
+        if self.header_rows:
+            for row in self.header_rows:
+                formatted_lines.append(self._format_row(row, colwidths))
+            formatted_lines.append(sep_row)
+
+        # Add column names
+        formatted_lines.append(self._format_row(lines[0], colwidths))
+        formatted_lines.append(sep_row)
+
+        # Add data rows
+        for line in lines[1:]:
+            formatted_lines.append(self._format_row(line, colwidths))
+
+        # Add top and bottom separators
+        return [sep_row] + formatted_lines + [sep_row]
+
+    def _get_colwidths(self, lines):
+        return [max(len(str(row[i])) for row in lines) for i in range(len(lines[0]))]
+
+    def _make_sep_row(self, colwidths):
+        return ' '.join('=' * width for width in colwidths)
+
+    def _format_row(self, row, colwidths):
+        return ' '.join(str(item).ljust(width) for item, width in zip(row, colwidths))
+
\end{lstlisting}

\end{goldpatchbox}

\end{document}